%% file: paper_freeE_imp.tex
\documentclass[aps,prb,twocolumn,showpacs,amsmath,amssymb,showpacs]{revtex4-2}
\usepackage{graphics,graphicx}
\usepackage{color}
\usepackage{amssymb,amsmath}
\usepackage{mathtools}
\usepackage{bookmath}
\usepackage{bm}
\usepackage{feynmf}
\renewcommand{\Re}{{\rm Re\,}}
\renewcommand{\Im}{{\rm Im\,}}
\newcommand{\whA}{{\widehat A}}

\def\Zc{Z_{\rm charge}}
\def\Zs{\overset{\scriptscriptstyle\leftrightarrow}{Z}_{\rm spin}}
\begin{document}
\unitlength = 1mm
%~~~~~~~~~~~~~~~~~~~~~~~~~~~~~~~~~~~~~~~~~~~~~~~~~
\title{Free Energy of Non-uniform Disordered Superconductors} 

\author{Matthias Eschrig}
\email[Contact author: ]{matthias.eschrig@uni-greifswald.de}
\affiliation{Institute of Physics, University of Greifswald, 17489 Greifswald, Germany} 

\author{Anton B. Vorontsov}
\email[Contact author: ]{anton.vorontsov@montana.edu}
\affiliation{Department of Physics, Montana State University, Bozeman, Montana 59717, USA}

\date{August 19, 2026}

\begin{abstract} 
We extend the Luttinger-Ward free energy functional to disordered superconductors and superfluids with arbitrary scattering
mechanisms, including both impurity disorder and the presence of interfaces.  
The disorder is
taken into account within the self-consistent $t$-matrix approximation, thus allowing for arbitrary impurity scattering
strengths. It is shown that both the interface and the impurity scattering self-energy appear in the functional only implicitly, through
self-consistently determined fermionic propagators. The free energy functional is formulated in terms of a generalized
integral in the complex energy plane, which encompasses formulations both in terms of retarded/advanced propagators and in terms of
Matsubara Green's functions by appropriately choosing the integration path. 
It can be applied, e.g., to spatially
non-uniform and hybrid systems, triplet and other unconventional condensates, 
%strong-coupling superconductors, 
and
strongly-correlated Fermi liquids. 
A particularly useful formulation in terms of the quasiclassical propagators is applied to unconventional non-uniform
singlet and triplet superconductors and superfluids in an external Zeeman magnetic field. 
\end{abstract}
\maketitle
%~~~~~~~~~~~~~~~~~~~~~~~~~~~~~~~~~~~~~~~~~~~~~~~~~~~~~~~~~~~~~~~~~~~~~~~~~~~~~
%~~~~~~~~~~~~~~~~~~~~~~~~~~~~~~~~~~~~~~~~~~~~~~~~~~~~~~~~~~~~~~~~~~~~~~~~~~~~~

\section{Introduction}

Calculating the free energy of a superconducting system is a fundamental thermodynamic problem of wide-ranging
importance, from accessing thermodynamic parameters to evaluation of the relative stability of phases. 
This is particularly important in multi-component
systems, non-uniform superconducting states, or in superconducting hybrid structures. The most general form follows
within a many-body Green's function formalism from the Luttinger-Ward functional based on its various extremal %, which has various important extremal
properties. Realistic materials contain boundaries, dislocation, lattice imperfections and impurities, leading to the necessity to incorporate
disorder in the many-body description. 
Uncorrelated impurities can be treated very effectively using Abrikosov's theory of superconducting alloys,\cite{AGDbook}
while interface effects can be incorporated through the boundary conditions \cite{Nagato1996,Nagato1998a,Eschrig2009}. 

The standard theory of superconductivity starts with a Fermi-liquid normal state combined with the Nambu-Gor'kov matrix
Green's function formalism \cite{Nambu1960,Gorkov1958,Gorkov1959}. A systematic expansion in terms of the ratio 
$s \sim T_c/ T_F, \hbar q /p_F, \hbar \omega /E_F$ between low-energy
and high-energy ranges in (quasi)-momentum-energy space allows for an asymptotically correct approximation in terms of
an expansion in the small parameter $s \ll 1$ \cite{Serene1983}. The low-energy range includes a narrow shell around the
Fermi surface (width $\hbar q \ll p_F$, where $p_F$ is the Fermi momentum) and low excitation energies ($\hbar \omega \ll E_F$, where $E_F$ is the  Fermi energy); 
the high-energy range is the complement of the low-energy range. 
%\ME{This definition includes a formal low-energy
%cut-off $sE_F \ll \Lambda \ll E_F$, which - via the renormalized, i.e. measurable, Fermi velocity $v_F$ - defines a corresponding 
%cut-off in momentum $\Delta p = \Lambda/v_F$ around the Fermi surface. 
%}
To formally separate low and high energy ranges one introduces a technical cut-off parameter 
$sE_F \ll \Lambda \ll E_F$, which also defines a corresponding 
cut-off in momentum $\Delta p = \Lambda/v_F$ around the Fermi surface via the renormalized, i.e. measurable, Fermi
velocity $v_F$.
Three aspects arise in such an expansion \cite{Serene1983}: first, all self-energies arising in a
formal Feynman diagrammatic expansion \cite{Mahan} can be classified according to their power in $s$; second, in leading
order in the expansion parameter, the propagator is given by its quasiclassical approximation, which represents only the
low-energy excitations (quasiparticles); and third, band structure, interactions, self-energies, and couplings to
external fields are renormalized by high-energy excitations, thus giving rise to their respective quasiparticle
equivalents. The quasiparticle renormalization factor is eliminated from the low-energy theory by incorporating it into the effective interactions and potentials.
To leading order in $s$, the theory is independent of the precise choice of the low-energy cut-off
$\Lambda$, which, when it explicitly appears, has to be eliminated by appropriate renormalization schemes in favor of observable
quantities.  
The basic quantities in this framework are the low-energy propagators and the high-energy
effective interactions and potentials. 
While to leading order in $s$ the propagators are sharply peaked around the Fermi surface at low excitation energies, 
the interactions and potentials are nearly independent of energy and temperature across the low-energy range; their external momenta can therefore be restricted to the
Fermi momenta.
Furthermore, to leading order, these effective interactions and potentials are unaffected by the superconducting transition and temperature variations.

In the following, we will assume this description in terms of low-energy propagators when discussing the expansion of
the free energy in terms of Feynman diagrams. We will concentrate on weak-coupling superconductors in this work.
The corresponding theory including electron-phonon coupling leads to
Eliashberg theory of superconductivity \cite{Eliashberg1960}, as Migdal's approximation
\cite{Migdal1958} applies in leading order in $s$. The matrix structure of the propagator accounts for the intrinsic degrees of freedom (e.g.,
particle-hole and spin), and its dependence on relative momentum and center-of-mass coordinate accounts for the
extrinsic degrees of freedom. Self-consistency for the self-energies, which are functionals of the low-energy
propagators, ensures conservation laws in the inhomogeneous state; for energy-flow conservation see \eg \cite{Richard2016}.

In order to incorporate disorder, we will use all diagrams in the disorder-averaged formalism of Abrikosov to leading
order in the small parameter $s$. This can be shown to be equivalent to the $t$-matrix approximation
\cite{Balatsky2006}. %\cite{T-matrix in FLT}. 
Keeping all the corresponding diagrams allows for a wide range of applications, including Born and unitary limits
within Eilenberger theory, as well as the diffusive limit within Usadel theory. The goal is to derive a free energy
functional that is correct within $t$-matrix approximation and therefore consistent with the standard theory of
superconductivity. 

To effectively treat inhomogeneous superconducting systems the theory is formulated in terms of Wigner coordinates
that include ``center-of-mass'' $\vR = (\vr_1 + \vr_2)/2$ and relative $\vr = \vr_1 - \vr_2$
coordinates, followed by a %consecutive 
Fourier transform $\vr \to \vp$, 
so that two-point functions such as propagators and self-energies are represented as $ f(\vr_1,\vr_2) \sim f_W(\vp, \vR)$. 
In addition, we present a derivation of the free energy that uses path integration in the complex energy ($z$) plane. 
This serves several purposes.
First, it directly relates the Matsubara energy functional to the propagators on the real energy axis, which in turn are
directly related to the spectral properties of the system. 
This unifies and expands several approaches existing in the literature. 
Second, it allows for a natural transition between the full
(Gor'kov) propagators and the quasiclassical propagators.  The transition from the full propagators to quasiclassical
propagators within the Luttinger-Ward functional is not trivial due to the presence of the log-operator, especially in
non-uniform superconductors. To accommodate
the corresponding ``$\xi$-integration'' one must resort to ``tricks'', such as differentiating the functional with
respect to some parameter (like the interaction or the self-energy), and adding ``counter terms''.\cite{Serene1983} 
%% 
%%  \AV{We are not using symmetric contour at all, but instead carefully eliminate possible divergencies by considering
% asymptotic behavior ... }   
%% We use a different approach, relying on a
%% special symmetric choice of the integration path in the complex energy ($z$) plane. This choice allows effectively to
%% take care of boundary terms and eliminates convergence issues for large $|z|$. 
%%
With the contour integration approach, the natural differentiation parameter is the energy itself, 
which effectively allows one to take care of boundary terms and eliminate convergence issues for large $|z|$ 
when calculating the free energy relative to the normal state's value. 
The formalism we present allows for a
formulation on both the real or the imaginary energy axes by simply deforming the integration path. 
Since the main derivation `procedure' involves manipulation of scalar energy parameter and no other characteristics, 
the free energy expression derived in this paper is quite general; it applies to 
superconductors with an arbitrary order parameter matrix, band structure and impurity scattering. 
Finally, it allows for a numerically very stable and efficient implementation of free energy calculations.

The quasiclassical free energy for superconductors with impurities has been derived in several limits previously, 
starting with the original work by Eilenberger,\cite{Eilenberger:1968wb}
which treated impurities in the Born limit. 
More recent calculations of the quasiclassical free energy starting from the Luttinger-Ward functional 
include the differentiation ``trick'' with respect to a scaling parameter of the self-energies 
in non-uniform films of spin-triplet p-wave \He.\cite{Vorontsov2003} 
A variation of the same trick was used to derive the Eilenberger free energy functional for spin-triplet correlations 
\cite{Virtanen2020} with impurities in the diffuse Born limit. 
To treat the thermodynamics of \He\ in aerogel, the authors of \cite{Ali2011} 
introduced a continuous Matsubara energy variable to do the $\xi$-integration; they  
used random uniform impurities as an external potential that,
after configuration averaging, appeared as a $t$-matrix self-energy in the propagators. 

We present the derivation of the general free energy functional in Section~\ref{sec:formalism}, 
%with the main results for free energy in terms of full Green's functions appearing in Eq.~\eqref{FEfullG}
with the main results for the free energy in terms of full Green's functions and energy-independent mean fields appearing in Eq.~\eqref{FINAL},
and in terms of quasiclassical propagators in Eqs.~\eqref{QC1} and \eqref{eq:QCM}. 
To demonstrate the variety of possible applications of the presented free energy functional, 
we consider several well-known systems that are considered in the subsequent sections: 
a clean uniform singlet $s$-wave superconductor in Section~\ref{sec:cleanS};
unitary and Born impurities in a disordered unconventional $d$-wave superconductor within a Zeeman magnetic field in Section~\ref{sec:disorderedD};
and
triplet superfluid \He\ with interface scattering, including interactions with magnetic field and Fermi-liquid effects, in Section~\ref{sec:triplet}.

%~~~~~~~~~~~~~~~~~~~~~~~~~~~~~~~~~~~~~~~~~~~~~~~~~~~~~~~~~~~~~~~~~~~~~~~~~~~~~~~~~~~~~~~~~~~~~~~~~

\section{General Formalism} 
\label{sec:formalism} 
%\include{Matthias_FreeEnergy_file}

%\ME{Use \textbackslash ME and \textbackslash AV for comments in blue and red.}

%\AV{
	%several things to pay attention to: 
	%(1) do trace first, inside the energy- or $dz$-integral;
	%(2) Check what happens when we use $\hat\Sigma^{mf}(\vare_m)$, - can we do that for generality;
%}

We start with the thermodynamic functional that was first introduced by Luttinger and Ward \cite{Luttinger1960}, 
and described in details in the standard treatises \cite{AGDbook,Serene1983}. At its stationary points it describes the grand canonical thermodynamic potential. In terms of the 4$\times $4 Nambu-Gor'kov matrix Green's function $\widehat G(i\varepsilon_n; \vp, \vR)$, 
self-energy $\widehat \Sigma(i\varepsilon_n; \vp, \vR)$, and external potentials $\widehat U(\vp,\vR)$
(e.g. an external magnetic field), it is given by
%\footnote{compared to \cite{Ali2011} we do not include impurities as external potential, but rather through a self-energy generating functional.}
%
\begin{widetext}
\begin{eqnarray}
\label{KBF}
\Omega [\widehat G, \widehat \Sigma; \widehat U ] = \Omega_0[\widehat U] 
-{\rm Tr} \Big\{ \widehat \Sigma \otimes \widehat G + \ln_\otimes (-\widehat G_0^{-1} + \widehat U + \widehat \Sigma ) 
-\ln_\otimes (-\widehat G_0^{-1} +\widehat U  )
\Big\}
+ \Phi[\widehat G] \,.
\end{eqnarray}
\end{widetext}
The last term, $\Phi[\widehat G]$, is a functional of the full Green's function only.
Sometimes the $\Omega$ functional is also referred to as Kadanoff-Baym functional
\cite{Kotliar2006}; for history and more details see \cite{martin2016interacting}. The self energy $\widehat \Sigma $ is treated as independent variable, and becomes a functional of the full Green's function $\widehat G$ under the condition of stationarity of the thermodynamic functional with respect to $\widehat G$ (see below).
The bare Green's function for a system without disorder is %, non-interacting state is 
\be
\widehat G_0^{-1}(i\varepsilon_n; \vp,\vR) = i \varepsilon_n \widehat \tau_3 - \xi_\vp \widehat 1 \,.
\label{G0} 
\ee
In order to conform with the notation convention of
quasiclassical Green's functions \cite{Serene1983}, we have included a factor $\widehat \tau_3 $ in $\widehat G_0^{-1}$, 
and similarly a factor $\widehat \tau_3 $ on the left in $\widehat G$ and on the right in $\widehat \Sigma $.
Note that the band structure entering the bare Green's function is already renormalized by ``high-energy'' processes. As
mentioned in the Introduction, ``high-energy'' quasiparticle weight factors also renormalize all external ``legs'' 
of potentials $\widehat U$, self energies $\widehat \Sigma $, and of internal interaction vertices appearing in the expressions of the self-energies, as
%and the quasiparticle weight was absorbed into the self-energy. 
%The external interaction $\widehat U$'s vertex also has the ``high-energy'' renormalization incorporated, as 
described in the review by Serene and Rainer \cite{Serene1983}. 
High-energy contributions to the free energy are part of the reference value 
$\Omega_0[\widehat U] = \Omega^{(0)}_0 +\Omega^{(1)}_0[\widehat U] + \Omega^{(2)}_0[\widehat U] + \dots$, 
where the linear-in-$\widehat U$ part $\Omega^{(1)}_0[\widehat U]$ determines the equilibrium quantities in the unperturbed
normal state, 
and the quadratic-in-$\widehat U$ part $\Omega^{(2)}_0[\widehat U] = \frac12 \widehat U \widehat\chi^{high} \widehat U$ yields
the high-energy contributions to the susceptibilities. 
Effectively, the difference inside the trace in Eq.~\eqref{KBF} includes only the low-energy quantities. 

In our treatment, we do not include randomly distributed impurities as external potential, as in Ref.~\cite{Ali2011}, but rather through a self-energy generating functional for configuration-averaged Green's functions and related self energies within self-consistent $t$-matrix approximation.
The electronic interactions that lead to superconductivity, Fermi-liquid corrections, and (configuration-averaged) interactions with 
impurities will be generated through the functional $\Phi[\whG]$, as we describe below.

Electromagnetic potentials $(\phi, \vA)$ are treated in leading order in an expansion in $s$ as part of $\widehat U$, as described in Appendix~\ref{EM}, leading to $\widehat U=\widehat U^{0}_a+\widehat{U}_{\rm EM}$ where 
$\widehat U_a^{0}$ refers to (fixed) non-electromagnetic external potentials. %in the absence of electromagnetic terms. 
We split the electromagnetic term into a part due to the (fixed) external electromagnetic fields, $\widehat U^{\rm ext}_{\rm EM}$, and a part generated by electromagnetic sources inside the material (e.g., supercurrents), $\widehat U^{\rm ind}_{\rm EM}$. 
The latter part becomes a functional of $\widehat G$ when coupled with Maxwell's equations, and the first part will be included into the external potentials, $\widehat U_a=\widehat U_a^0+\widehat U^{\rm ext}_{\rm EM}$. Furthermore, the total thermodynamic functional contains in this case two contributions, $\Omega_{\rm tot}=\Omega+\Omega_{\rm EM}$, 
where $\Omega$ is for the superconducting material
and its coupling to electromagnetic potentials, 
while $\Omega_{\rm EM}$ depends only on $\widehat{U}_{\rm EM}$ and describes the electromagnetic field energy. The total functional $\Omega_{\rm tot}$ is a functional of $\widehat G$, $\widehat \Sigma$, and the electromagnetic potentials.
It is stationary against variation with respect to the electromagnetic potentials, and the condition of stationarity leads to the (electrostatic) Maxwell's equations, which couple the electromagnetic potentials to the full Green's function $\widehat G$ via the electromagnetic source terms generated by $\Omega $. Note that, as a result, 
%Notice, however, that 
at the stationary point, the potential $\widehat U=\widehat U_a+\widehat{U}^{\rm ind}_{\rm EM}[\widehat G]$ is not necessarily 
the applied external potential $\widehat U_a$, as electromagnetic fields are modified in the superconducting state.
We concentrate in the following on the material part $\Omega$, 
and refer for details of how the electromagnetic potentials are incorporated to Appendix~\ref{EM}. 
As shown there, the consistent implementation of the vector potential in the material part $\Omega$ to leading order in
$s$ allows the use of the convolution product in Wigner representation in terms of canonical momenta.

The convolution product in Wigner representation is defined as
\begin{widetext}
\begin{eqnarray}
[\widehat A\otimes \widehat B](i\varepsilon_n; \vp,\vR) = e^{-\frac{i}{2} (\nabla_{\vp_1}\nabla_{\vR_2}-\nabla_{\vp_2}\nabla_{\vR_1})} \widehat A(i\varepsilon_n;\vp_1,\vR_1) \widehat B(i\varepsilon_n;\vp_2,\vR_2) \Big|_{\substack{\vp_1=\vp_2=\vp\\ \vR_1=\vR_2=\vR}}
\end{eqnarray}
\end{widetext}
which involves a matrix multiplication as well.
Some properties of the convolution products are listed in Appendix \ref{OTIMES}.
The inverse as well as the logarithm are understood to be taken with respect to the $\otimes $-product, i.e. $\whA^{-1}$ is the solution of $\whA^{-1} \otimes \whA=\widehat 1$, and 
$\ln_\otimes (\widehat 1+\whA)=\whA-\frac{1}{2}\whA\otimes \whA+\frac{1}{3} \whA\otimes
\whA\otimes \whA+\cdots $.
The trace is defined as
\begin{eqnarray}
{\rm Tr}\{ \cdots \} = 
T\sum_{\varepsilon_n} \; \frac{1}{2} \mbox{Tr}_\otimes \left\{ \cdots \right\},
\end{eqnarray}
where fermionic Matsubara energies are $\varepsilon_n=(2n+1)\pi k_\sm{B}T$ and
\begin{eqnarray}
{\rm Tr}_\otimes \{ \cdots \} = 
\int d^3\vR \int \frac{d^3\vp}{(2\pi)^3}
\; {\rm tr}_4 \{\cdots \},
\end{eqnarray}
where ${\rm tr_4}$ is a trace over combined particle-hole and spin space.
%\footnote{ Note that we have included here in the trace a factor of $\frac{1}{2}$.}
We use an extended zone scheme, where 
%the $i^{\rm th}$ energy band refers to the $i^{\rm th}$ order Brillouin zone, and
integrating $\vp$ over all Brillouin zones is equivalent to integrating over the first Brillouin zone combined with a
sum over band indices (this allows for simplifications in the notation).  We will also assume that only a finite number
of energy bands need to be taken into account, by using renormalized effective interaction vertices that include all
high-energy contributions from the neglected bands (see the review by Serene and Rainer \cite{Serene1983} for
details). This assumption avoids convergence issues associated with the sum over an infinite number of energy bands.
The low-energy range includes only a momentum-shell around a finite number of distinct Fermi surface pockets. 
Using the following definition of the functional derivative, 
\begin{eqnarray}
%\frac{\delta X[\widehat A] }{\delta \widehat A } &=& \widehat B [\widehat A]
\delta X [\widehat A]= {\rm Tr} \left\{ 
\frac{\delta X[\widehat A] }{\delta \widehat A } \otimes \delta \widehat A \right\} \,,
\label{func_deriv}
\end{eqnarray}
the stationary point of functional (\ref{KBF}) is given by the Dyson equation, 
\begin{eqnarray}
-\frac{\delta \Omega }{\delta \widehat \Sigma } 
= \widehat G - [ \widehat G_0^{-1} - \widehat U - \widehat \Sigma ]^{-1} = \widehat 0
\label{Dyson}
\end{eqnarray}
and the self-consistency equation for skeleton self-energy diagrams $\widehat \Sigma$, 
generated from the functional $\Phi [\widehat G ]$, is %\cite{AGDbook}
\begin{eqnarray}
-\frac{\delta \Omega }{\delta \widehat G } = 
\widehat \Sigma [\widehat G ] - \frac{\delta \Phi [\widehat G ] }{\delta \widehat G } 
= \widehat 0 \,.
\label{SelfE}
\end{eqnarray}
The stationary properties of the functional and the generating functional $\Phi$ 
were first emphasized in \cite{Baym1962,Klein1961}, see also \cite{AGDbook}.
The definition of the functional derivative, Eq.~\eqref{func_deriv}, relies on the invariance of the trace under cyclic permutations of the operators; note that this trace involves a full spatial integration $\int d^3\vR$, see Appendix
\ref{OTIMES}.

Note that the energy dependence of the Green's functions and self energies appearing in the Dyson equation and the self-consistency equation can be analytically extended from the Matsubara energies $z=i\varepsilon_n$ to complex energies $z$, in both the upper and lower complex half plane. For $z$-values infinitesimally close
to the real axis, their solutions yield the retarded and advanced 
Green functions, respectively, depending on the sign of the imaginary part of $z$.

Our goal is to evaluate the extremum value of the free energy functional \eqref{KBF} 
by using the self-consistently determined Green's function \eqref{Dyson} and the self-energies \eqref{SelfE}. 

\begin{figure*}[t]
\includegraphics[width = 0.31\linewidth]{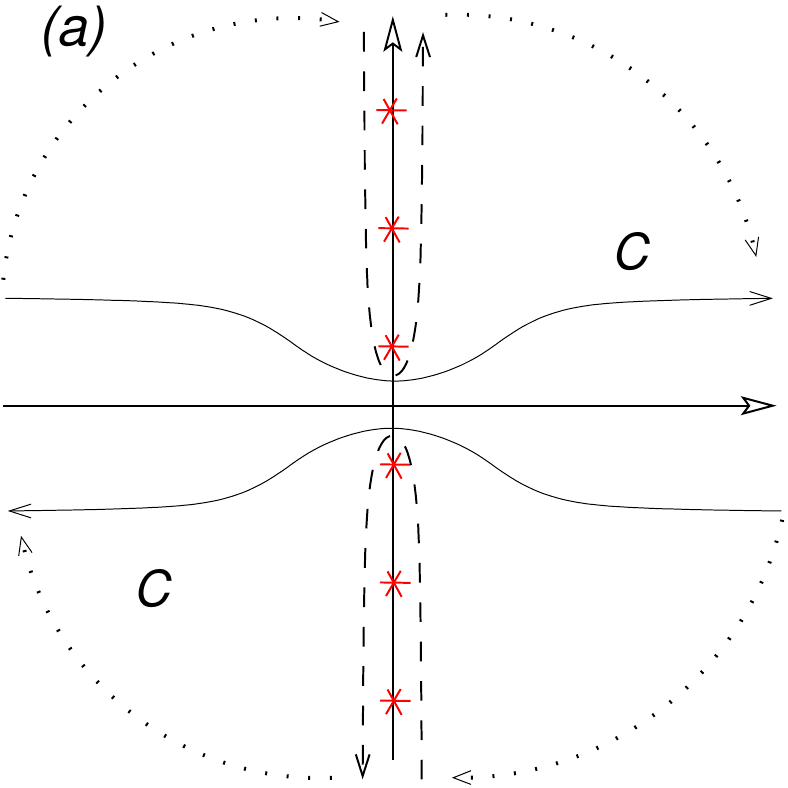}
\hfill
\raisebox{3mm}{\includegraphics[width = 0.31\linewidth]{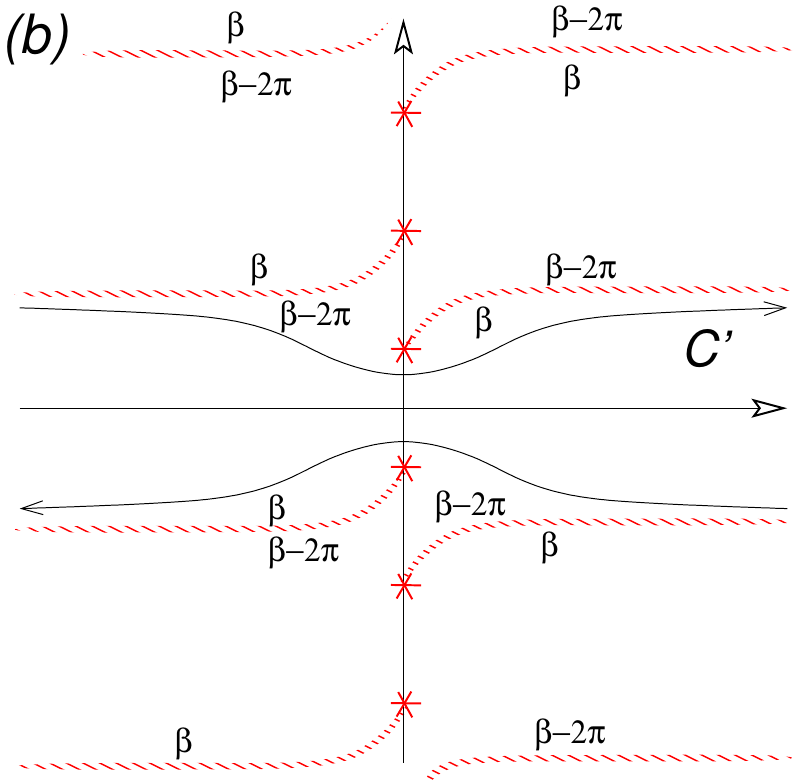}}
\hfill
\includegraphics[width = 0.31\linewidth]{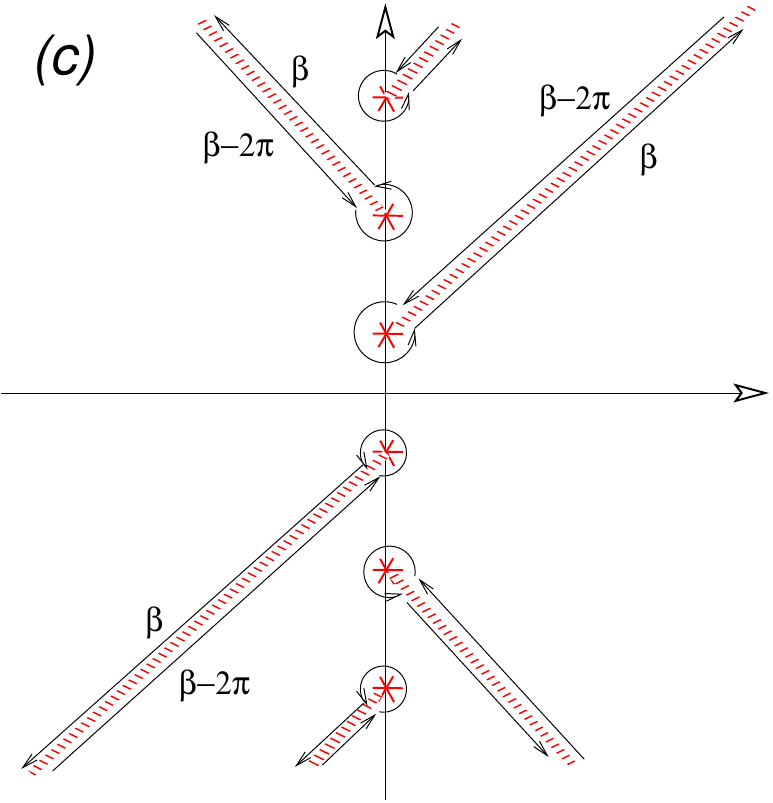}
\caption{ \label{fig:contours} 
Various integration paths of the free energy functional. 
(a) The original dashed-line contour around the Matsubara poles 
$z^*=i\varepsilon_n = i \pi T (1+2n)$ (stars) of the $\tanh(z/2T)$ function can 
be modified to run differently in the complex energy plane. 
This contour can be used up to Eq.~(\ref{CC}). 
(b) 
After integration by parts in Eqs.~(\ref{PI}-\ref{PIomega}), we generate a non-analytic function $\log(\cosh z)$, and 
the contour has to be completely in the analyticity domain of 
this function. The branch cut of the $\log w$-function is quite arbitrary, and we show it for the case 
$(\Re w, \Im w) = (t \cos\beta, t\sin \beta)$ with $t\in[0,\infty)$ and $\beta=3\pi/4$.  
%in this case is chosen as the $Im w = -Re w$ in the second quadrant of complex $w$. 
The argument of $w=\cosh z$ is shown on the edges of the branch cuts. 
%Integration by parts is done on each part separately, and 
This contour should be used in Eq.~(\ref{CCC}). 
(c) 
We can deform the branch cuts, and wrap the integration contour along them. When the branch cuts
are along the $y$-axis this integration results in free energy expression Eq.~(\ref{eq:QCM}). 
%\AV{(the straight lines in $z$ correspond to spiral branch cut of $\log w$ in $w$-plane!)}
}
\end{figure*}

We use the fact that we can write the sum over the Matsubara frequencies
as a contour integral around the Matsubara energy poles of the hyperbolic tangent, 
\begin{eqnarray}
\label{formula}
T\sum_{\varepsilon_n} \widehat F(i\varepsilon_n)= \oint \frac{dz}{4\pi i} \widehat F(z) \tanh \frac{z}{2T}
\end{eqnarray}
and we can modify the integration contour 
under the assumption that the contribution from remote circle arcs at $|z|\to \infty $ vanishes, 
\begin{eqnarray}
\label{cond}
\lim_{|z|\to \infty} z\widehat F(z) = \widehat 0 \; ,
\end{eqnarray}
and the new contour does not encompass any new poles from $\widehat F(z)$. % relative to the contour is the same. 
An example of an `unfolding' is shown in Fig. \ref{fig:contours}(a). 
The resulting contour is rather general, but has two branches, one in the upper and one
in the lower half plane. The first branch starts in the second quadrant far away
from the coordinate origin (at $-\infty + i\Gamma $),
passes the imaginary axis between the zeroth and the first positive Matsubara
frequency, and continues into the first quadrant to values far away from the
coordinate origin (to $\infty + i\Gamma $). 
The second branch is equal to the first one 'inverted' %mirrored 
at the coordinate origin. 
Apart from this the contour can be arbitrarily chosen. 
The symmetrically inverted contour 
%also minimizes numerical calculations as it makes possible to use symmetries of the propagators and self-energies, and 
replaces Eq.~\eqref{cond} with a weaker condition for the vanishing contributions from the remote circle arcs 
\begin{eqnarray}
\label{cond_symm}
\lim_{|z|\to \infty} z \left[ \widehat F(z) + \widehat F(-z) \right] = \widehat 0 \; .
\end{eqnarray}
The most efficient numerical evaluation is for contours that are mirror-symmetric around both the real and imaginary
$z$-axes, as then full advantage of all propagators' symmetries can be taken. 

% \AV{We never use the inverted-symmetric path it in this version} 
% Note that the inverted image property of the two branches 
% %(in fact only mirroring of the remote circle arcs is needed) 
% allows to formulate a weaker 
% %As a result of this mirror image property of the two branches, 
% condition for the vanishing contributions from the remote circle arcs: 
% \begin{eqnarray}
% \label{cond1}
% \lim_{|z|\to \infty} |z| \left[\widehat F(z)+\widehat F(-z) \right] = \widehat 0 \; ,
% \end{eqnarray}
% %\AV{(do we actually use it below in my new formulation? - yes, but somewhere at the end, around eq(33).)} 
% which exploits the fact that $\tanh(z/2T)$ is odd in $z$. 
% We will make clear when we use this condition. 
% %This condition is used later
% %to cast the self-energy generating functional $\Phi[\whG]$ for mean-fields into
% %a form with contour integration along real energy. 

%%%%%%%%%%%%%%%%%%%%%%%%%%%%%%%%%%%%%%%%%%%%%%%%%%%%%%%%%%%%%%%%%%%%%%%%%%%%%%%%%
%%%%%%%%%%%%%%%%%%%%%%%%%%%%%%%%%%%%%%%%%%%%%%%%%%%%%%%%%%%%%%%%%%%%%%%%%%%%%%%%%

We define a modified trace with integration along the contour, 
\begin{eqnarray}
{\rm Tr_C}\{ \cdots \} = 
\oint \frac{dz}{4\pi i} \;
\frac{1}{2}\mbox{Tr}_\otimes \left\{
\cdots  \right\}
\,.
%\int d^3\vR \int \frac{d^3\vp}{(2\pi)^3}
%\; \frac{1}{2} {\rm tr}_4 
\end{eqnarray}

%\AV{ My changes really START } 

\begin{widetext}
To modify the integration contour we first look at the asymptotic $z$-behavior of various terms in the integrand of Eq.~\eqref{KBF}. 
Leaving the self-energy generating term $\Phi[\whG]$ aside for the moment, we consider just the difference term 
\begin{eqnarray}
\label{int}
%\Omega_0[\widehat G, \widehat \Sigma ] - \Omega_0=
\Del\Omega_0 \equiv 
-{\rm Tr} \Big\{ 
\widehat \Sigma \otimes \widehat G +
\ln_\otimes (-\widehat G_0^{-1} + \widehat U + \widehat \Sigma ) 
-\ln_\otimes (-\widehat G_0^{-1} + \widehat U ) 
\Big\} \; . \quad
\end{eqnarray}
Applying the matrix operations as described in appendix \ref{OA}, and using 
the expansion $\whG = \whG_0 + \whG_0 \otimes (\widehat U + \widehat \Sigma) \otimes \whG_0 + \dots$ 
with $\whG_0(z) = (\widehat\tau_3 z - \xi_\vp)^{-1}$, the asymptotics 
of the function inside the trace
$\mbox{Tr}_\otimes $ 
for large $|z|$ 
%the leads to a term $-\widehat G_0\otimes \Sigma $ which in combination with the first term 
in Eq.~(\ref{int}) is 
\begin{eqnarray}
%\mbox{Tr}_\otimes \left\{ \lim_{|z|\to \infty} \left(\widehat G(z)-\widehat G_0(z)\right) \otimes \widehat \Sigma \right\} 
%= 
%\mbox{Tr}_\otimes \left\{ \lim_{|z|\to \infty} \left(\widehat G_0(z) \otimes \widehat
%\Sigma \right)^2 \right\} \sim O\left( \frac{1}{z^2}\right)
-\lim_{|z|\to \infty} 
{\rm Tr}_\otimes \left\{ \widehat \Sigma \otimes \widehat G +
\ln_\otimes (-\widehat G_0^{-1} + \widehat U + \widehat \Sigma ) 
-\ln_\otimes (-\widehat G_0^{-1} + \widehat U ) 
\right\} 
 = 
-{\rm Tr}_\otimes \left\{ \frac 12 \left[ \whG_0(z) \widehat\Sigma(z) \right]^2  
+ \cO \left( \whG_0\right)^3  \right\}
\label{int_asympt}
\end{eqnarray}
\end{widetext}
This quantity has to decay faster than $1/|z|$ for large $|z|$ in order to satisfy condition (\ref{cond}) for arbitrary contour deformation, 
and even for convergence of the original Matsubara sum. 
At the first glance, it would seem that this is indeed the case, since in this limit $\whG_0 \to O(1/z)$  
(see Appendix \ref{ASYMPTG} for more details) and the Green's function appears at least twice in the expression; however
such a conclusion does not, in fact, hold because of the $\xi$-integration inside ${\rm Tr}_\otimes$. 
Breaking up the self-energy into diagonal, $\widehat\nu \propto \widehat 1,\widehat \tau_3 $, and off-diagonal, $\whDelta\propto\widehat \tau_1, \widehat \tau_2 $, components in particle-hole space 
\begin{align}
\widehat \Sigma &= \begin{pmatrix} \nu & \Delta \\ \ul\Delta & \ul\nu \end{pmatrix} \equiv 
\widehat \nu + \whDelta\,,
%\\
%\widehat\nu &\propto \widehat 1,\widehat \tau_3\,,\quad 
%\whDelta\propto\widehat \tau_1, \widehat \tau_2 \nonumber
\end{align} 
one can see that the $[\whG_0 \widehat\Sigma]^2$ terms that involve matrix products $\widehat\nu \whDelta$ 
will disappear after performing the Nambu trace  
${\rm tr_4} \{ \whDelta \widehat\nu \} = 0$, while terms proportional to the square $\widehat\nu^2$ will vanish after
integration 
$$
\lim_{|z|\to \infty} \int_{-\infty}^{+\infty} d\xi \frac{1}{(\xi \pm z)^2} = 0 
$$
However, for the anomalous part of the self-energy,  
$\widehat\Delta$, the combination $[\whG_0(z) \whDelta]^2 \propto \whDelta^2 /(\xi^2 - z^2)$ 
and the $\xi$-integration in ${\rm Tr}_\otimes$ will produce %(remember $z$ has imaginary part) 
$$
\lim_{|z|\to \infty} \int_{-\infty}^{+\infty} d\xi \frac{\widehat\Delta^2}{\xi^2 -z^2} 
\sim \cO \left( \sgn(\Im{ z}) \frac{\whDelta^2(z)}{z}\right)
$$
Thus, the convergence of the ensuing Matsubara sum or $z$-integral depends on the high-energy properties 
of the off-diagonal self-energy. If the anomalous self-energy decays with $z$, 
$\whDelta(z\to \infty) \to 0$, then there is no issue. However, if the self-energy is non-decaying, say
$\whDelta(z)=const$, then there is a
log-divergence in $z$-summation that needs to be regulated. This is the familiar divergence from the mean-field BCS pairing. 
We show how to deal with these two cases by considering mean-field and impurity scattering self-energies. 

We need explicit forms for self-energies and for $\Phi[\widehat G]$. 
The general form of these functional is described in the review by Serene and Rainer \cite{Serene1983}. 
We will assume that both the self-energy and the generating functional are sums of two terms, 
an impurity term and mean-field energy-independent term. 
Thus, we have 
$$ \widehat \Sigma(z)= \widehat \Sigma_{imp}(z)+\widehat \Sigma^{mf}(z) $$ 
and
$$\Phi[\widehat G]=\Phi_{imp}[\widehat G]+\Phi^{mf}[\widehat G] \,.$$

\subsection{Mean-field self-energies}

%%%%%%%%%%%%%%%%%%%%%%%%%%%%%%%%%%%%%%%%%%%%%%%%%%%%%%%%%%%%
\begin{figure*}[t!]
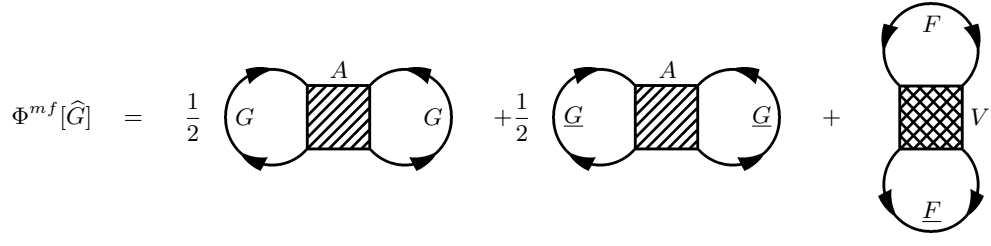

\input feynman_diagrams_mf
\caption{ \label{fig:mfPhi} 
	The free energy functional to generate mean field self-energies. We only include the weak-coupling diagrams here. }
\end{figure*}
%%%%%%%%%%%%%%%%%%%%%%%%%%%%%%%%%%%%%%%%%%%%%%%%%%%%%%%%%%%%

We explicitly write the mean-field %break the mean-field terms and propagators into diagonal and off-diagonal components, 
self-energy and propagator in the particle-hole space: 
\begin{equation}
\widehat \Sigma^{mf} = \left( 
\begin{array}{cc} 
\nu & \Delta \\ \ul\Delta & \ul\nu 
\end{array} 
\right) 
\;, \qquad
\widehat G = \left( 
\begin{array}{cc} 
G & F \\ \ul F & \ul G
\end{array} 
\right) \,,
\end{equation} 
The self-energies are self-consistently determined from the propagators: 
\begin{widetext}
\begin{align} \begin{split} 
\label{eq:mfsc}
& \nu( i\vare_n; \vp,\vR) = T \sum_{\vare_{n'}} \int \frac{d^3\vp'}{(2\pi)^3} W(i\vare_n, \vp; i\vare_{n'}, \vp') G(i\vare_{n'}; \vp', \vR) \\
%& \ul\nu(i\vare_n; \vp,\vR) = T \sum_{\vare_{n'}} \int \frac{d^3\vp'}{(2\pi)^3} A(i\vare_n, \vp; i\vare_{n'}, \vp') \ul G(i\vare_{n'}; \vp', \vR) 
%\\
& \Delta(i\vare_n; \vp,\vR) = T \sum_{\vare_{n'}} \int \frac{d^3\vp'}{(2\pi)^3} 
V(i\vare_n, \vp; i\vare_{n'}, \vp') F(i\vare_{n'}; \vp', \vR) \\
%& \ul\Delta(i\vare_n; \vp,\vR) = T \sum_{\vare_{n'}} \int \frac{d^3\vp'}{(2\pi)^3} 
%V(i\vare_n, \vp; i\vare_{n'}, \vp') \ul F(i\vare_{n'}; \vp', \vR) 
\end{split} \end{align} 
and the $\ul\nu$ and $\ul\Delta$ are related to these through the particle-hole symmetry, 
\eg $\ul \nu(z; \vp,\vR) = \nu(-z^*; -\vp, \vR)^*$.
Here we explicitly show only momentum and energy sums, while the spin degrees of freedom are implicitly included into 
the interaction vertices, 
$W(i\vare_n, \vp; i\vare_{n'}, \vp')$ and $V(i\vare_n, \vp; i\vare_{n'}, \vp')$, which we generalized to be energy-dependent and
analytic. 
The Fermi-liquid interactions, when confined to the low-energy domain and combined with the normal density of states at the Fermi level, $N_f$, 
are described by the dimensionless interaction traditionally denoted by $A$: 
$A(i\vare_n, \vp_f; i\vare_{n'}, \vp'_f) = N_f W(i\vare_n, \vp_f; i\vare_{n'}, \vp'_f)$, following notation in \cite{Eschrig2015a}.
Similarly, a dimensionless pairing interaction  $\lambda(i\vare_n, \vp_f; i\vare_{n'}, \vp'_f) = N_f V(i\vare_n, \vp_f; i\vare_{n'}, \vp'_f) $ is often introduced.
In the typical BCS approximation the off-diagonal interactions $V(i\vare_n, \vp, i\vare_{n'}, \vp')$ 
can be taken as energy independent, but in this case it is necessary to cut-off the $\vare_{n'}$-summation at some $\Lambda$ scale to avoid the
log-divergence. 
This cut-off energy, and the interaction $V$, can be removed from the theory. 
The final results are insensitive to the exact procedure for elimination of the log-divergence, and one can use the standard trick 
of eliminating them in favor of the superfluid/superconducting transition temperature in the self-consistency
equation\cite{Serene1983}. 
Here we show that these quantities also do not appear in the free
energy functional expressed in terms of the self-consistently determined $\widehat\Sigma^{mf}(z)$ and $\whG(z)$. 

The self-consistency equations (\ref{eq:mfsc}) are generated from the functional in
Fig.~\ref{fig:mfPhi}, where each fermionic loop comes with its momentum integration and
Matsubara sum, and the $\otimes$-product of propagators. The appropriate symmetries of the vertices as well
as the spin summations are assumed. 
Using the definition of the self-energies (\ref{eq:mfsc}) we can eliminate one of the propagators and the interaction
vertex in favor of
self-energy and write this functional as: 
\begin{eqnarray}
\Phi^{mf}[\widehat G] = \frac{1}{2}{\rm Tr } \left\{ \widehat \Sigma^{mf} \otimes \widehat G
\right\}
\end{eqnarray}

We now combine the mean-field generating functional with Eq.~\eqref{int}: 
\begin{eqnarray}
\label{int_mf}
\Del\Omega_0 + \Phi^{mf}[\widehat G] = 
-{\rm Tr} \Big\{ 
\widehat \Sigma \otimes \widehat G - \frac12 \widehat \Sigma^{mf} \otimes \widehat G
+\ln_\otimes (-\widehat G_0^{-1} + \widehat U + \widehat \Sigma ) 
-\ln_\otimes (-\widehat G_0^{-1} + \widehat U ) 
\Big\} \; . \quad
\end{eqnarray}
If one proceeds with the asymptotic analysis in an analogous way as in the derivation of Eq.~\eqref{int_asympt},
one obtains for the limit of large $|z|$:  
\begin{eqnarray}
-\lim_{|z|\to \infty} 
{\rm Tr}_\otimes \left\{ \whG_0(z) \widehat\Sigma^{mf}(z) \whG_0(z) \widehat\Sigma_{imp}(z) + \frac 12 \left[ \whG_0(z) \widehat\Sigma_{imp}(z) \right]^2  
+ \cO \left( \frac{1}{z^3} \right)  \right\} 
\propto \cO \left( \frac{1}{z^2} \right)  
\end{eqnarray}
as all the potentially problematic $\whDelta_{mf}^2/(\xi^2-z^2)$ terms disappear, and the remaining terms vanish either due to
${\rm tr_4}$-summation or $\xi$-integration. The impurity self-energy correction in $t$-matrix approximation 
only contributes in higher order, $\widehat\Delta_{imp}(z) \propto {\widehat\tau_{1,2}}/{z}$. 
Consequently, even for the case of energy-independent $\widehat\Sigma^{mf}$ we can perform the $\xi$-integration, and then sum over the
Matsubara energies, and the free energy expression will converge. 

Since the integrand in Eq.~(\ref{int_mf}) fulfills the condition (\ref{cond}), we can transition to a contour
integration: 
\begin{eqnarray}
\label{CC}
\Del\Omega_0  &+& \Phi^{mf}[\widehat G] = 
\\ &&
= -{\rm Tr_C}\left\{ 
\tanh \frac{z}{2T} \; \left(
\widehat \Sigma(z) \otimes \widehat G - \frac12 \widehat \Sigma^{mf}(z) \otimes \widehat G
+ \ln_\otimes [-\widehat G_0^{-1}(z)  + \widehat U + \widehat \Sigma (z) ]
-\ln_\otimes [-\widehat G_0^{-1}(z)  + \widehat U]
\right)
\right\}
\nonumber
\end{eqnarray}
We still have to add the impurity self-energy generating functional $\Phi_{imp}[\widehat G]$ 
to obtain the correct self-energy, but before we do so we eliminate 
the logarithms in this expression. 
To do that we perform a partial integration using 
\begin{eqnarray}
%\int_{z_0}^{z} 
\tanh \frac{z}{2T} dz = d\left[ 2T \ln \left( 2\cosh \frac{z}{2T} \right) \right] 
% + {\rm const.}
\label{PI}
\end{eqnarray}
where a factor of 2 inside the logarithm is introduced as a convention to obtain the correct energy factor at low $T$. %\cite{} 
As the function $\ln (\cosh z)$ is non-analytic, we make sure the contour $C$ 
does not cross its branch cuts. 

%where the integral is over any specified contour from $z_0$ to $z$.
%The (possibly temperature dependent) constant 
%is irrelevant according what was said above. 
%The next observation is that the contour integral over the right hand side
%in Eq. (\ref{CC}) {\it without} the $\tanh $-factor is zero.
%This is a consequence of the analytic properties of the Green functions.
%
%Note in passing that
%\begin{eqnarray}
%2T \ln \left( 2\cosh \frac{\epsilon }{2T} \right) 
%&=& \epsilon  + 2T \ln (1+e^{-\frac{\epsilon }{T}} ) \\
%&=& |\epsilon | + 2T \ln (1+e^{-\frac{|\epsilon |}{T}} ) 
%\; .
%\end{eqnarray}
%
%Partial integration gives %for the logarithmic part of the free energy
%If the main part of the contour is below the first branch cut, 
For the choice of the contour 
as shown in Fig.~\ref{fig:contours}(b), partial integration gives 
\begin{eqnarray}
\Del\Omega_0 + \Phi^{mf}[\widehat G] &=& 
+{\rm Tr_C} \Big\{ 
2T \ln \left( 2\cosh \frac{z}{2T} \right)  
\Big[
\partial_z \left( \widehat \Sigma (z) \otimes \widehat G(z) \right) 
- \frac12 \partial_z \left( \widehat \Sigma^{mf} (z) \otimes \widehat G(z) \right) +
\Big.  \; \Big. \label{PIomega} \\ &&\Big.  \Big.
\left( \widehat G_0^{-1}(z) - \widehat U - \widehat \Sigma (z) \right)^{-1}  \otimes
	\partial_z \left( \widehat G_0^{-1}(z) - \widehat U - \widehat \Sigma (z) \right)
-\left( \widehat G_0^{-1}(z) - \widehat U \right)^{-1} \otimes \partial_z \left( \widehat G_0^{-1}(z) - \widehat U\right)
\Big]
\Big\} \; .
\nonumber
\end{eqnarray}
Note that we used the invariance of the trace under cyclic permutations
in order to write the derivative of the logarithm term (see appendix \ref{OA}).
We can now again use the fact that the Green's functions are
solutions of the Dyson equation. 
Since  
$\left( \whG_0^{-1}(z) - \widehat U - \widehat \Sigma (z)\right)^{-1} = \whG(z)$, and 
$\partial_z \widehat G_0^{-1}(z) = \widehat \tau_3 $
we obtain
\begin{eqnarray}
\label{CCC}
\Del\Omega_0 + \Phi^{mf}[\widehat G] = 
+{\rm Tr_C} \Big\{ 
2T \ln \left( 2\cosh \frac{z}{2T} \right)  && 
\Big[ \widehat \tau_3 \left( \whG(z)- [\whG^{-1}_0(z) -\widehat U]^{-1} \right)
\\ \nonumber 
&& + \widehat \Sigma (z) \otimes \partial_z\whG(z) 
- \frac12 \partial_z \left( \widehat \Sigma^{mf} (z) \otimes \widehat G(z) \right) 
\Big]
\Big\} \; .
\end{eqnarray}

\subsection{Impurity self-energy in $t$-matrix approximation}

We now add the impurity-generating functional $\Phi_{imp}[\whG]$. Note that in Eq.~\eqref{CCC} 
the $\widehat \Sigma (z) \otimes \partial_z\whG(z)$-term still contains the full self-energy 
$\widehat \Sigma (z) = \widehat \Sigma_{imp} (z) +\widehat \Sigma^{mf} (z)$. 
We now show that for impurities in the $t$-matrix
approximation considerable simplifications occur, and 
$\Phi_{imp}[\whG]$ cancels the $\widehat \Sigma_{imp} (z) \otimes \partial_z\whG(z)$ contribution. 
The $t$-matrix equation reads
\begin{eqnarray}
\widehat t (z;\vp_1,\vp_2,\vR)= \widehat v(\vp_1,\vp_2) + \int \frac{d^3\vp' }{(2\pi)^3}\widehat v(\vp_1,\vp') \otimes \widehat G (z;\vp',\vR)\otimes \widehat t(z;\vp',\vp_2,\vR) .
\label{TM}
\end{eqnarray}
Here, $\widehat v(\vp_1,\vp_2)$ is the scattering potential for a single impurity, 
and the impurity self-energy is 
$$
\widehat \Sigma_{imp} (z;\vp,\vR)=c_{imp}\widehat t(z;\vp,\vp,\vR) \,,
$$ 
where $c_{imp}$ is the impurity concentration.
%We assume that the constant $\widehat v_0=\int \frac{d^3\vp}{(2\pi)^3} \widehat v(\vp,\vp)$ 
%is included into $\xi_{\vp }$ as it merely renormalizes the electrochemical potential. 
%Thus, without loss of generality we assume that $\mbox{Tr}_\otimes \left\{\widehat v(\vp,\vp)\right\}=\widehat 0$.

%%%%%%%%%%%%%%%%%%%%%%%%%%%%%%%%%%%%%%%%%%%%%%%%%%%%%%%%%%%%
\begin{figure*}[t]
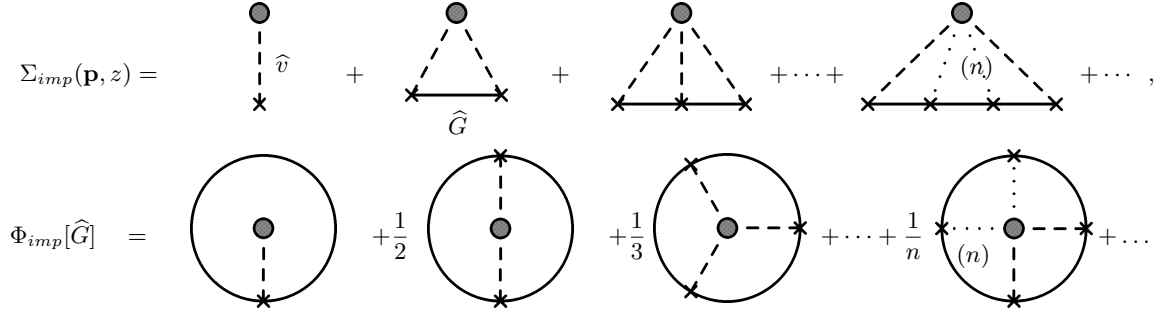

\input feynman_diagrams_imp
\caption{ \label{fig:iPhi} 
	The diagrams for $t$-matrix and the generating functional}
\end{figure*}
%%%%%%%%%%%%%%%%%%%%%%%%%%%%%%%%%%%%%%%%%%%%%%%%%%%%%%%%%%%%

The $\Phi$-functional consists of all `wheel'-diagrams with corresponding
symmetry pre-factors, shown in Fig. \ref{fig:iPhi}. 
The symmetry factor for the ring diagram with
$n$ impurity interaction lines is ${1}/{n}$. There is an overall
sign of $-1$ due to a fermion loop. Thus we have
%\begin{eqnarray}
%\Phi_{imp}[\widehat G] = -c_{imp}{\rm Tr } \left\{\sum_{n=1}^{\infty} \frac{1}{n} (\widehat G \otimes \widehat v )^n \right\} 
%\nonumber \\
%=
%c{\rm Tr } \ln \Big\{[\widehat 1 - \widehat G \otimes \widehat v]^{-1} \Big\}.
%\end{eqnarray}
\begin{align} \begin{split} %\begin{eqnarray}
\Phi_{imp}[\widehat G] &= -c_{imp}{\rm Tr } \left(
\widehat G (i\varepsilon_n;\vp,\vR)\otimes \widehat v(\vp,\vp)+ 
\frac{1}{2}\int \frac{d^3\vp'}{(2\pi)^3}
\widehat G (i\varepsilon_n;\vp,\vR)\otimes \widehat v(\vp,\vp') \otimes \widehat G
(i\varepsilon_n;\vp',\vR)\otimes \widehat v(\vp',\vp) \right. 
\\ %\nonumber \\
& 
\left. + \frac{1}{3}\int \frac{d^3\vp'}{(2\pi)^3} \frac{d^3\vp''}{(2\pi)^3}
\widehat G (i\varepsilon_n;\vp,\vR)\otimes \widehat v(\vp,\vp') \otimes \widehat G (i\varepsilon_n;\vp',\vR)\otimes \widehat v(\vp',\vp'')\otimes \widehat G (i\varepsilon_n;\vp'',\vR)\otimes \widehat v(\vp'',\vp) + \cdots \right),
\end{split} \end{align} % \end{eqnarray}
Note that the Tr operation includes an integration over $d^3\vp/(2\pi)^3$, which will be important for the following discussion.
We now replace the sum over the Matsubara frequencies by a contour integral, noticing that the boundary term vanishes as the asymptotic behavior of the leading term, which gives rise to $\mbox{Tr}_\otimes \left\{ z\widehat G (z;\vp,\vR)\otimes \widehat v(\vp,\vp) \right\}$, for large $|z|$ is 
$ \mbox{Tr}_\otimes \left\{ \widehat \tau_3 \widehat v(\vp,\vp) \right\}
\propto \mbox{tr}_4 \left\{ \widehat \tau_3 \widehat v(\vp,\vp) \right\}
=\widehat 0$,
which follows from $\widehat v(\vp,\vp)=v(\vp,\vp)\widehat 1$.
Thus, Eq.~\eqref{cond} is again fulfilled.
Performing subsequently a partial integration using Eq.~\eqref{PI}, and noticing that boundary terms again vanish (for the same reason as above), we obtain
\begin{align} \begin{split} %\begin{eqnarray}
\Phi_{imp}[\widehat G] = & - c_{imp} {\rm Tr_C } \bigg\{ %\left\{
2T \ln \left( 2\cosh \frac{z}{2T} \right)
\partial_z \bigg( %\left(
%\right. \right. 
\\ & \left. \left. 
\widehat G (z;\vp,\vR)\otimes \widehat v(\vp,\vp)
+ \frac{1}{2}\int \frac{d^3\vp'}{(2\pi)^3}
\widehat G (z;\vp,\vR)\otimes \widehat v(\vp,\vp') \otimes \widehat G (z;\vp',\vR)\otimes \widehat v(\vp',\vp) \right. \right. 
\\ & \left. \left. 
+ \frac{1}{3} \int \frac{d^3\vp'}{(2\pi)^3} \frac{d^3\vp''}{(2\pi)^3}
\widehat G (z;\vp,\vR)\otimes \widehat v(\vp,\vp') \otimes \widehat G
(z;\vp',\vR)\otimes \widehat v(\vp',\vp'')\otimes \widehat G (z;\vp'',\vR)\otimes
\widehat v(\vp'',\vp) + \cdots \right) \right\}.
\end{split} \end{align} % \end{eqnarray}

For each term with $n$ functions $\widehat G$ there will be $n$ derivative terms.
However, we can use the invariance of the trace $\mbox{Tr}_\otimes $ with respect to cyclic permutation to bring the terms $\partial_z \widehat G$ all to the rightmost position in each term. Having done this, all $n$ derivative terms become identical expressions, and the corresponding factor $n$ cancels the corresponding pre-factor $1/n$ in front of each original term.
This leads to
\begin{align} \begin{split} %\begin{eqnarray}
\Phi_{imp}[\widehat G] =& -c_{imp}{\rm Tr_C } \left\{
2T \ln \left( 2\cosh \frac{z}{2T} \right) \;
\left(
\widehat v(\vp,\vp) +
\int \frac{d^3\vp'}{(2\pi)^3} \widehat v(\vp,\vp')\otimes
\widehat G (z;\vp',\vR)\otimes \widehat v(\vp',\vp) + \right. \right. \nonumber \\
&\left. \left. 
\int \frac{d^3\vp'}{(2\pi)^3} \frac{d^3\vp''}{(2\pi)^3} \widehat v(\vp,\vp')\otimes
\widehat G (z;\vp',\vR)\otimes \widehat v(\vp',\vp'') \otimes \widehat G (z;\vp'',\vR)\otimes \widehat v(\vp'',\vp) + \cdots \right) 
\otimes \partial_z\widehat G (z;\vp,\vR) \right\}
\end{split} \end{align} % \end{eqnarray}
The term in the round brackets is, however, nothing else than the expression obtained for the $\widehat t$-matrix by the method of successive substitution in Eq.~\eqref{TM}, taken for coinciding arguments $\vp_1=\vp_2=\vp$. Thus, we obtain
\begin{eqnarray}
\Phi_{imp}[\widehat G] &=& -c_{imp}{\rm Tr_C } \left\{
2T \ln \left( 2\cosh \frac{z}{2T} \right) \;
\widehat t(z;\vp,\vp,\vR) \otimes \partial_z\widehat G (z;\vp,\vR) \right\}.
\end{eqnarray}
We now substitute $\widehat \Sigma_{imp} (z;\vp,\vR)=c_{imp}\widehat t(z;\vp,\vp,\vR)$ and obtain
\begin{eqnarray}
\Phi_{imp}[\widehat G] &=& -{\rm Tr_C } \left\{
2T \ln \left( 2\cosh \frac{z}{2T} \right) \;
\widehat \Sigma_{imp}(z;\vp,\vR) \otimes \partial_z\widehat G (z;\vp,\vR) \right\}.
\end{eqnarray}
This term precisely cancels the corresponding term in Eq.~\eqref{CCC} 
involving $\widehat \Sigma_{imp}(z) \otimes \partial_z\widehat G (z) $.
Note that this conclusion holds for any generally anisotropic scattering potential $v(\vp,\vp')$ for elastic scattering from normal impurities. It can also straightforwardly be generalized to the case of various types of normal impurities with different scattering characteristics and concentrations.

Thus, we can omit the impurity self energy in Eq. (\ref{CCC}) if we
also omit at the same time the impurity contribution to the $\Phi$-functional, giving the full free energy functional: 
\begin{eqnarray}
\Omega[\widehat \Sigma_{imp} &+& \widehat \Sigma^{mf}, \whG; \widehat U]  - \Omega_0[\widehat U] = 
\label{FEfullG}
\\
&=&
{\rm Tr_C} \Big\{ 
2T \ln \left( 2\cosh \frac{z}{2T} \right)  
\; 
\left[ \widehat \tau_3 \left( \whG(z)- [\whG^{-1}_0(z) -\widehat U]^{-1} \right)
+\widehat \Sigma^{mf} (z) \otimes \partial_z\whG(z) 
- \frac12 \partial_z \left( \widehat \Sigma^{mf} (z) \otimes \whG(z) \right)
\right]
\Big\} 
\nonumber 
\end{eqnarray}
This is one of the central results of this publication. 

In the following, to simplify expressions, we assume the mean-fields are energy-independent: 
$$ \widehat\Sigma^{mf}(z) = \mbox{const} $$
In this case the free energy further reduces to

\begin{eqnarray}
\label{FINAL}
\Omega - \Omega_0&=&
+{\rm Tr_C} \Big\{ 
2T \ln \left( 2\cosh \frac{z}{2T} \right)  
\; 
\left[\widehat \tau_3 \left( \widehat G(z)-[\widehat G^{-1}_0(z) -\widehat U]^{-1} \right)
+\frac{1}{2}\widehat \Sigma^{mf} \otimes \partial_z \widehat G(z)\right]
%-\frac{1}{2} \left( \tanh \frac{z}{2T} \right) 
%\left(\widehat \Sigma^{mf} \widehat G(z) - \widehat \Sigma_N^{mf} \widehat G_N(z) \right)
%\widehat \Sigma^{mf} \widehat G(z)
\Big\} \; .
\end{eqnarray}
This is the final expression appropriate for BCS superconductors and superfluids. 
The impurity self energies and the \textit{}$\Phi_{imp}$ functional drop out of the expression, 
but the Green functions $\whG$ are the fully impurity averaged ones, and impurity self-consistency is performed in the
``background'' together with self-consistency on $\widehat\Sigma^{mf}$. 

%\AV{
%At this point the high-energy divergencies after $\xi$- and $z$-integrations 
%in  
%$\widehat \tau_3 \left( \widehat G(z)-[\widehat G_0^{-1}(z) - \widehat U]^{-1} \right)$ term 
%is canceled by the $\frac12 \widehat \Sigma^{mf} \partial_z \widehat G(z)$ term. 
%So we can take the mean-field self-energies to be energy-independent all the way to $|z|=\infty$. 
%Direct calculation using asymptotyc expansion 
%Eq.\eqref{asymptG} with energy-independent mean fields gives for the RHS of \eqref{FINAL}
%% \be
%% \int^\Lambda dz \int d\xi \; {\rm tr}_4 \left[ 
%% 	2T \ln \left( 2\cosh \frac{z}{2T} \right)  \; 
%% \widehat\tau_3 \frac{\whDelta^2 }{ (\xi^2 - z^2)(z\widehat\tau_3 - \xi) } \right]
%% \propto \whDelta^2 \ln \Lambda \;,
%% \ee
%% is regulated by the $\whDelta^2$ term from $\frac12 \widehat \Sigma^{mf} \partial_z
%% \widehat G(z)$. 
%\be
%\int d\xi \; {\rm tr}_4 \left\{ 
%	2T \ln \left( 2\cosh \frac{z}{2T} \right)  \; \widehat \tau_3 \left[ 
%	\frac12 \widehat\nu \whG_0^2 + \whG_0^3 \widehat\nu\widehat U + \whG_0^3 \widehat U \whDelta + \frac12 \whG_0
%	\whDelta \whG_0 + O(z^{-4}) \right] 
%	\right\}
%\ee
%The first three terms in $[\dots]$ are zero after $\xi$-integration, due to poles of $\vG_0$-powers being on one side of
%complex plane. The other term is zero due to ${\rm tr}_4 (\widehat\tau_3 \whDelta) = 0 $. 
%The remainder term, after integration over $\xi$ in the highenergy limit becomes at most $O(z^{-2})$ - convergent. 
%}

Eq. (\ref{FINAL}) can then directly be $\xi_p$-integrated to obtain the
quasiclassical version, since the integrand is a sufficiently narrow function
of $\xi_p$ due to the derivative $\partial_z$, and is decaying quickly enough for $z\to \pm \infty $. 
It is common to relate the thermodynamic potential
to that of the normal state, $\Omega_N[\widehat U_a]$, for the same external conditions, such as temperature, and external potentials $\widehat U_a$ (here we take potentials in the normal state to be identical to the applied ones).
The correct thermodynamic potential after $\xi_p$-integration reads
%\begin{widetext}
\begin{eqnarray}
\label{QC}
	\Omega[\widehat U] - \Omega_N [\widehat U_a]&=&
	\Omega_0[\widehat U] - \Omega_0 [\widehat U_a] 
	\\ 
&+&{\rm Tr'_C} \left\{ 
2T \ln \left( 2\cosh \frac{z}{2T} \right)  
\; 
\left[
\widehat \tau_3 \Big( \widehat g(z)-\widehat g_N(z) \Big)
%\Big.\nonumber \\ &&\Big. \qquad
%-\frac{1}{2} \left( \tanh \frac{z}{2T} \right) 
+\frac{1}{2}\Big(\widehat \sigma^{mf} \partial_z\widehat g(z) - \widehat \sigma_N^{mf} \partial_z\widehat g_N(z) \Big)
\right]
\right\}
\nonumber
\end{eqnarray}
\end{widetext}
%\AV{isn't $\widehat \sigma_N^{mf}=0$ here if we took care of high-energy contributions in normal state already?} 
with
\begin{eqnarray}
{\rm Tr'_C}\{ \cdots \} = 
\oint \frac{dz}{4\pi i} \; 
\int d^3R \; 
\frac{1}{2} {\rm tr}_4 \; N_f \langle \cdots \rangle_\sm{FS},
\end{eqnarray}
where $\langle \cdots \rangle_\sm{FS}$ denotes Fermi surface averaging, and $N_f$ is the normal density of states at
Fermi level per one spin projection.

The (energy) $z$-derivatives of the quasiclassical propagator may be inconvenient, and moreover they produce higher
singularities at the poles and branching points of the propagator. These points require special care in the contour
integration, 
%by going around them, and not easy to implement 
especially when integrating along 
retarded and advanced branches, with complex energies infinitesimally close to the real-energy axis. 
To eliminate these derivatives we perform a partial integration of the two last terms inside % $+\frac12( \dots)$-term in 
Eq.~\eqref{QC}, % to get an alternative form of free energy, 
in which case, however, care must be taken of the resulting boundary terms.
At high energies the asymptotic expressions for the quasiclassical 
Green's function are (Appendix \ref{ASYMPTG}), 
\be
%\widehat f(z) &=& \pm i \pi \frac{\widehat \Delta^{mf} }{z} + \cdots
%\widehat g (z)-\widehat f(z) &= & \mp i \pi \widehat \tau_3 
%\left( \widehat 1- \frac{[\widehat \Delta^{mf}]^2}{2z^2} + \cdots \right)
\widehat g (z) = 
\sgn(\Im{ z}) \; i \pi \left[ - \widehat \tau_3 
%\left( \widehat 1- \frac{[\whDelta^{mf}]^2}{2z^2} \right)
+ \frac{\whDelta^{mf}}{z} \right] + O\left(\frac{1}{z^2} \right) 
\label{HEqcg}
\ee
%where the upper sign is for the retarded branch. 
We place here a superscript, $\Delta^{mf}$, in ordere to emphasize that this is a mean-field order parameter, 
and the impurity contribution enters only in higher order terms. 
As a result, the correct form of the thermodynamic potential after partial integration in $z$ reads
\begin{widetext}
\begin{eqnarray}
\label{QC1}
\Omega[\widehat U] - \Omega_N [\widehat U_a] &=& \Omega_0[\widehat U] - \Omega_0 [\widehat U_a] 
+{\rm Tr'_C} \Big\{ 
2T \ln \left( 2\cosh \frac{z}{2T} \right)  
\; 
\widehat \tau_3 \Big( \widehat g(z)-\widehat g_N(z) \Big)
\Big.\nonumber \\ &&\Big. \qquad
-\frac{1}{2} \left( \tanh \frac{z}{2T} \right) 
\Big(\widehat \sigma^{mf} \widehat g(z) - \widehat \sigma_N^{mf} \widehat g_N(z) \Big)
\Big\}  
%\nonumber \\ &&
+ \frac{1}{2}N_f \int d^3R \frac{1}{2} {\rm tr}_4 \Big\langle [\whDelta^{mf}]^2 \Big\rangle_\sm{FS} 
.
\end{eqnarray}
\end{widetext}
The extra term is the boundary term appearing during the partial integration. 
Notice that for the real-axis contour $C$ consisting of retarded and advanced branches, 
the propagator difference $\whg^R(\vare) - \whg^A(\vare)$ is proportional to the density of states, 
and thus the free energy is directly related to spectral properties of the system. 

Finally, we present the free energy functional in terms of the Matsubara energies. 
For that we modify the branch cuts as indicated in Fig.~\ref{fig:contours}(c), and align
them along the imaginary $z=i\lambda$ axis (in the figure we show them at an angle, for
clarity, and the contour is inversion-symmetric by construction). 
Then we wrap the integration contour around branch cuts, and use it in
Eq.~(\ref{QC}). Function $2T\ln(2\cosh z/2T)$ has discontinuity $4\pi i T $ across each cut, resulting in 
\begin{widetext}
\be
\Omega[\widehat U] - \Omega_N [\widehat U_a] = \Omega_0[\widehat U] - \Omega_0 [\widehat U_a] 
+ {\rm Tr'} \left\{ 
\int\limits_{\vare_m}^{\infty \, {\rm sgn}(\vare_m)} 
i d\lambda \; \widehat{\tau}_3 \left[ \whg(i\lambda;\vp,\vR) - \whg_N(i\lambda) \right] 
-  \frac12 \whs^{mf}(\vp,\vR) \whg(i\vare_m; \vp, \vR) \right\}
\;,\qquad 
\label{eq:QCM}
\ee
\end{widetext}
with 
\be
{\rm Tr'}\{ \cdots \} = 
T \sum_{\vare_m} \; 
\int d^3R \; 
\frac{1}{2} {\rm tr}_4 \; N_f \langle \cdots \rangle_\sm{FS},
\ee
and where we eliminated the upper integration limit and the last term, $\whs^{mf}_N \whg_N$, 
by using $\sum_{\vare_n} \whg_N(i\vare_n) = 0$. 
Notice that the choice of integration when branch cuts are running along the $y$-axis is fundamentally the same 
as the original integration contour around the Matsubara energies, and no `unfolding' of integration contour occurs; 
in this case the discussion of $|z|\to \infty$ behavior of various functions
is still relevant if we want to extend the $\lambda$-integration to infinity to avoid dealing with explicit cutoffs. 

Functional \eqref{eq:QCM} is identical in the form to that obtained in \cite{Ali2011}, which was derived by introducing
external impurity potential and performing impurity configuration average, rather than through the generating functional as in
this paper. 

It is convenient to make both terms under the trace in \eqref{eq:QCM} to converge independently. 
To do that, we remove from the propagator in the last term the high-energy asymptotics Eq.\eqref{HEqcg} due to the off-diagonal (anomalous) 
mean field self-energy, 
\begin{widetext}
\begin{align*}
\whs^{mf} \whg(i\vare_m) 
& = \whs^{mf} \left( \whg(i\vare_m) - \pi \frac{\whDelta^{mf}}{|\vare_m|} \right)
+ \frac{\pi}{|\vare_m|} \whs^{mf} \whDelta^{mf} 
\\ & =
\whs^{mf} \left( \whg(i\vare_m) - \pi \frac{\whDelta^{mf}}{|\vare_m|} \right)
+ \pi  \int\limits_{\vare_m}^{\infty{\rm sgn}\vare_m} \frac{d\lambda \, {\rm sgn} \vare_m }{\lambda^2} 
\; \whs^{mf} \whDelta^{mf} 
\,.
\end{align*}
We add the last term to the $\lambda$-integral term in \eqref{eq:QCM}, and use the fact that 
in the last term  the diagonal mean fields drop out after particle-hole summation, 
${\rm tr_4} [ \whs^{mf} \whDelta^{mf}] = {\rm tr_4} [\whDelta^{mf}]^2 $, 
\bea 
\label{eq:QCMmodf}
\Omega[\widehat U] - \Omega_N [\widehat U_a] &=& \Omega_0[\widehat U] - \Omega_0 [\widehat U_a] 
%\Omega - \Omega_N = 
\\
&+& {\rm Tr'} \left\{ 
\int\limits_{\vare_m}^{\infty \, {\rm sgn}(\vare_m)} 
d\lambda \left( \widehat{\tau}_3 \, i \left[ \whg(i\lambda) - \whg_N(i\lambda) \right]  
- \pi \frac{ {\rm sgn} \vare_m }{2\lambda^2} [\whDelta^{mf}]^2\right) 
-  \frac12 \whs^{mf} \left( \whg(i\vare_m) - \pi \frac{\whDelta^{mf}}{|\vare_m|} \right)\right\}
\;.\qquad 
\nonumber
\eea
%\end{widetext}
We suppress the $(\vp,\vR)$-arguments in $\whg,\whs^{mf},\whDelta^{mf}$ for easy readablilty. 
In this expression the integration and summation are explicitly convergent because the asymptotic behavior of the integrand
becomes $\cO(\lambda^{-3})$, after particle-hole space summation of the asymptotic form of the propagator, 
Eq. \eqref{asymptQCG} with $z= i \lambda$. 
The last term without the integration can be simplified further by using the self-consistency equation on mean
fields. An example of this procedure in shown later in \eqref{eq:FE}. 

In the examples below, and as often is assumed in the literature when neglecting Meissner screening effects, 
we assume the Zeeman applied field to be unaffected by the superconducting state, $\widehat U = \widehat U_a$ and thus 
$\Omega_0[\widehat U] - \Omega_0 [\widehat U_a] =0$.

% ABV - we already discussed the large-z asymptotics above 
%Note that the integral in the trace in Eq. (\ref{QC1}) is convergent for large $|z|$.
%This can be seen for example in quasiclassical approximation for a singlet
%superconductor.
%For large $\epsilon={\rm Re} z$ (and fixed finite Im $z$) the first summand in Eq. (\ref{QC1}) goes like
%$|\epsilon |\widehat \tau_3 (\widehat g(z)-\widehat g_N(z))$, which in the superconducting state has the
%leading term $\mp i\pi |\Delta|^2/2|\epsilon|$ (depending on the branch in the
%upper ($-$)or lower ($+$) half plane), which cancels against
%the large $|\epsilon |$ asymptotics of the second term $-\frac{1}{2} 
%{\rm sign} (\epsilon )\widehat \Delta  \widehat f (z)$, using that $\widehat f(z) \sim
%\pm i\pi \widehat \Delta / \epsilon $, and that $\widehat \Delta \widehat \Delta = -|\Delta|^2$. 
%

\section{Clean uniform singlet superconductor} 
\label{sec:cleanS} 

As an initial demonstration for the use of these formulas, 
we consider a uniform clean singlet superconductor, with only off-diagonal mean-fields (energy-independent) 
\begin{align*}
\widehat\sigma^{mf}(z;\vp) = \whDelta_\vp = \begin{pmatrix} 0 & \Delta_\vp i\sigma_2 \\ \Delta_\vp^* i\sigma_2 & 0 \end{pmatrix}
%\\
\qquad\mbox{and}\qquad
\whg_u(z;\vp) = -\pi \frac{z \widehat\tau_3 - \whDelta_\vp}{\sqrt{|\Delta_\vp|^2 - z^2}} \,,
\end{align*}
An intermediate result after taking ${\rm tr_4}$ and the continuous `Matsubara' energy integral $z=i\lambda$ in
Eq.~\eqref{eq:QCM}, is the free energy density 
$\Del\Omega^{mf} = (\Omega - \Omega_N)/V$  
%\begin{widetext}
$$
\Del\Omega^{mf} =  - 2\pi T \sum_{\vare_m} N_f \left\langle 
\sqrt{\vare_m^2 + |\Delta_\vp|^2} - |\vare_m| - \frac{|\Delta_\vp|^2}{2\sqrt{\vare_m^2 + |\Delta_\vp|^2}} \right\rangle_\sm{FS} \,,
$$
%\end{widetext}
showing how the last term eliminates the log-divergence of the Matsubara sum. 

If we instead consider real-energy integration, we take the $C$-contour to run just above and below the real energy axis
by setting $z=\vare \pm i 0$ and thus use retarded and advanced propagators, 
%\begin{widetext}
\begin{align*}
& \frac14 {\rm tr_4}\left[\widehat\tau_3(\whg^R - \whg^A)\right] 
=  -2 \pi i \; \Im \frac{\vare+i0}{\sqrt{|\Delta_\vp|^2 - (\vare+i0)^2}}
=  -2 \pi i \; \frac{|\vare|}{\sqrt{\vare^2 - |\Delta_\vp|^2}}\theta(\vare^2 - |\Delta_\vp|^2) 
\,,\qquad \\
& \frac14 {\rm tr_4}\left[\widehat\sigma^{mf}(\whg^R - \whg^A)\right] 
=  -2 \pi i \; \Im \frac{|\Delta_\vp|^2}{\sqrt{|\Delta_\vp|^2 - (\vare+i0)^2}}
=  -2 \pi i \; \frac{|\Delta_\vp|^2 \sgn \vare}{\sqrt{\vare^2 - |\Delta_\vp|^2}}\theta(\vare^2 - |\Delta_\vp|^2) 
\end{align*}
the intermediate result of Eq.~\eqref{QC1} is 
\begin{eqnarray}
&&\Del\Omega^{mf} = 
- N_f \langle |\Delta_\vp|^2 \rangle_\sm{FS} 
\\ \nonumber
&& - \left\langle  \int\limits_{-\infty}^{+\infty} d\vare \left\{
2T \ln \left( 2\cosh \frac{\vare}{2T} \right) 
\left[ \frac{|\vare| \; \theta(\vare^2 - |\Delta_\vp|^2) }{\sqrt{\vare^2 - |\Delta_\vp|^2}} - 1 \right]
-\tanh \frac{|\vare|}{2T} \frac{|\Delta_\vp|^2}{2} \frac{\theta(\vare^2 - |\Delta_\vp|^2) }{\sqrt{\vare^2 - |\Delta_\vp|^2}}
\right\} \right\rangle_\sm{FS}
\,.
\end{eqnarray}
%\end{widetext}

In the zero-temperature limit we recover the standard basic answer in both Matsubara and real-energy treatments: 
\be
\Del\Omega^{mf}(T=0) = - \onehalf N_f \langle |\Delta_\vp|^2 \rangle_\sm{FS}  \,.
\ee

%\AV{ My main modification ENDS} 

\section{Disordered $D$-wave superconductor} 
\label{sec:disorderedD} 

For a more complicated application of these formula we consider a singlet superconductor 
in a Zeeman magnetic field. 
In the absence of Fermi-liquid effects, $\whs^{mf}=\whDelta$ is only the off-diagonal order parameter, and 
$\vB = \vH$ is externally applied field. 
The order parameter is $\Delta(\vp,\vR) = \Delta(\vR) \cY(\hvp)$, 
with a $d$-wave symmetry basis function $\cY(\hvp) = (2\hp_x \hp_y)$, 
and the amplitude being a function of position $\vR$ for a generally non-uniform superconducting state. 

%with Fermi-liquid corrections $\whs^{mf}=\whDelta(\vR,\hvp) + \vnu(\vR) \widehat{\vS}$ 
%and $\vH \to Z_0^a \vH = (1-A_0^a) \vH$. 
One solves the quasiclassical transport equation
in the presence of magnetic field, mean-field order parameter and impurities, 
%\begin{widetext}
\be
[ i\vare_m \widehat{\tau}_3 - \mu \vH \cdot \widehat{\vS} - 
\whDelta(\vp, \vR) - \whs^{imp}(\vare_m; \vR) , 
\whg(\vare_m; \vp, \vR)]
+ i\vv_f(\vp) \cdot \gradR \; \whg(\vare_m; \vp, \vR)  = 0 \,. 
\label{eq:eilZ}
\ee
%\end{widetext}
The Zeeman coupling to the field is through the electron's magnetic moment $\mu$, 
and the spin matrix in particle-hole-spin space is, 
\be
\widehat{\vS} =\left( \begin{array}{cc}
\vsigma & 0 \\ 0 & \vsigma^* \end{array} \right) \,.
\ee
The mean-field order parameter amplitude $\Delta(\vR)$ satisfies the self-consistency 
%\begin{widetext}
\be
\left\langle \cY(\hvp)^2 \right\rangle_\sm{FS} \Delta(\vR) \ln {T\over T_{c0}} = T \sum_{\vare_m} 
\left\langle  \cY(\hvp) \left(
f(\vare_m; \vp, \vR)
-\frac{\Delta(\vR) \cY(\hvp)}{|\vare_m|} \right) \right\rangle_\sm{FS} 
\label{eq:sc}
\ee
\end{widetext}
while impurity self-energies $\whs^{imp} = n_{imp} \widehat{t}$ are obtained by solving the $\widehat{t}$-matrix equation,
    	\be 
	\widehat{t}(\vare_m; \vR)
    	= u \widehat{1} + u N_f \langle \whg(\vare_m; \vp,\vR) \rangle_\sm{FS}
    	\; \widehat{t}(\vare_m; \vR) 
	\,.
    	\label{Eq:t-matrix}
    	\ee
Impurities are parametrized by the
impurity scattering rate $\gamma = \Gamma \sin^2\delta=1/2\tau_N$, 
expressed via the scattering phase shift, $\delta = \arctan (\pi u N_f)$ 
and the impurity concentration, $\Gamma=n_{imp}/\pi N_f$. 

In the absence of effects that mix orbital and spin spaces 
the two spin projections de-couple and one can solve for spin-up and spin-down propagators
and impurity self-energy separately. 

We use the free energy expression in the form of Eq.~(\ref{eq:QCMmodf}), 
where the last term is proportional to the right-hand side of the self-consistency relation Eq.~(\ref{eq:sc}). 
Taking the trace and using the self-consistency relation, the free energy density is explicitly given by 
\begin{widetext}
% \be
% \Del \Omega (\vR) =  
% \onehalf N_f \; T\sum_{\vare_m} {\rm tr_4} 
% \left\langle \int\limits_{\vare_m}^{\infty sign\vare_m} i\widehat{\tau_3} 
% \left[ \whg(i\lambda; \vp, \vR) - \whg_N(i\lambda; \vp, \vR) \right] d\lambda 
%  -\frac12 \whg(i\vare_m; \vp, \vR)\whs^{mf}(\vp, \vR) \right\rangle_\sm{FS} \,.
% \label{eq:feF}
% \ee
\be
\Del \Omega (\vR) =  
N_f \left\langle \cY^2(\hvp) \right\rangle |\Delta(\vR)|^2 \ln {T\over T_{c0}}  %- \frac{\vnu(\vR)^2}{A_0^a} 
+N_f \; 2T\sum_{\vare_m>0} \int\limits_{\vare_m}^{\infty } d\lambda 
\left\langle i [ g_\uparrow(i\lambda; \vp, \vR) + g_\downarrow(i\lambda; \vp, \vR) ] -2\pi 
+ \frac{\pi |\Delta(\vR)|^2 \cY(\hvp)^2}{\lambda^2} \right\rangle_\sm{FS} \,.
\label{eq:FE}
\ee
\end{widetext}

This expression is valid for a singlet superconductor in a Zeeman magnetic field 
for any (potential) impurity scattering. 

It is important to note, however, that the {\it density} of the free energy 
is not a uniquely defined quantity for non-uniform condensates, and it is in general different 
depending for different formulations of the quasiclassical free energy problem. 
It is the spatially integrated free energy that is a meaningful quantity. Any function that results into zero when integrated over the spatial coordinates can be added or subtracted to the free energy density without changing the free energy. Therefore caution is needed in trying to find a physical interpretation of free energy densities.
We will illustrate this in section \ref{sec:dwsc}, see figure \ref{fig:densF}. 

\subsection{Uniform superconductor in magnetic field} 
\label{sec:usc}

We first look at a uniform disordered superconductor. 
For simplicity we take a 2D superconductor with cylindrical Fermi surface, and the field is in-plane, 
so that one can neglect vortex effects. These conditions correspond to Pauli-limited superconductors where the main
pairbreaking occurs due to destruction of singlet Cooper pairs by flipping the spins.  
At low temperatures the transition into the normal state is of the first 
order, which is known as the Chandrasekhar-Clogston limit. \cite{Chandrasekhar1962,Clogston1962}

The free energy functional derived above can be used to find how the first order transition is modified by the impurities. 
A somewhat related functional was also used in our previous study of Pauli-limited critical field in dirty $d$-wave
superconductors. \cite{VorontsovAB:2008jl} 

\begin{figure}[t]
\centerline{\includegraphics[width=1.0\linewidth]{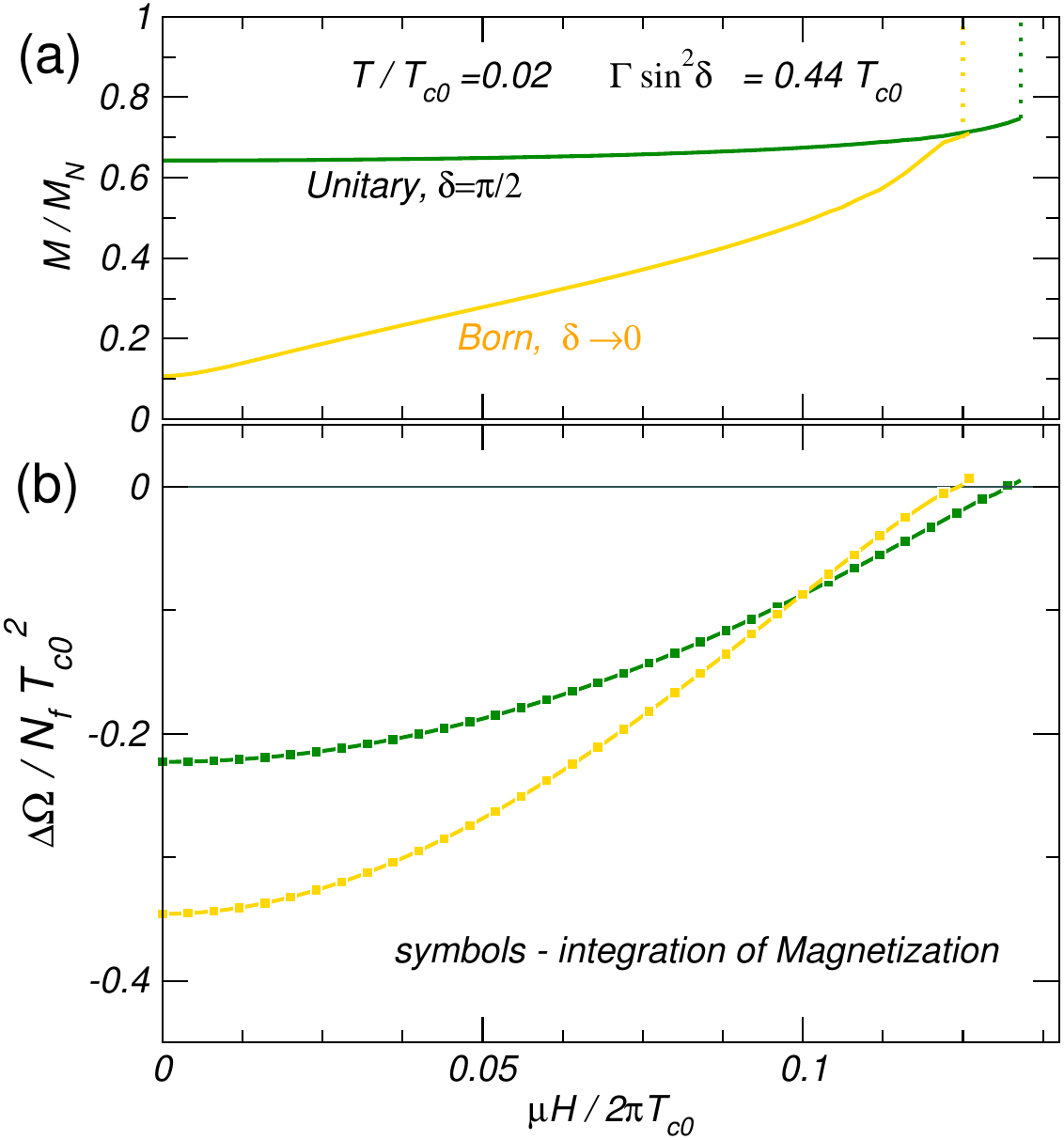}}
\caption{ \label{fig:FfromM}
Disordered uniform Pauli-limited $d$-wave superconductor at low temperatures, $T=0.02 T_{c0}$. 
The scattering rate $\gamma = \Gamma \sin^2\delta = 0.44 T_{c0}$ is the same for unitary ($\delta=\pi/2$) 
and Born ($\delta \to 0$) impurities. 
(a) Magnetization, and (b) free energy, as functions of (Zeeman) magnetic field. 
%The two scattering limits are significantly different, but both show first order phase transition into the normal state,
%where the magnetization experiences a discontinuous jump to its normal state value.  
%Even though the scattering rate $\gamma$ is the same, Born impurities result in a slightly lower critical field 
%compared to unitary impurities. 
The free energy in (b) is calculated in two ways: directly using Eq.~(\ref{eq:FE}) shown by the line, 
and by integration, Eq.~(\ref{eq:FfromM}), of the magnetization curves in (a), shown by the symbols. 
Both methods give the same result within numerical accuracy. 
}
\end{figure}

It is convenient to use the transition temperature without impurities, $T_{c0}$ as a reference value.
In this example we set the impurity scattering rate to be $\gamma = \Gamma \sin^2\delta = 0.44 T_{c0}$, 
which in zero field gives a suppression of the transition temperature, $T_c \approx 0.63 T_{c0}$,
according to the Abrikosov-Gor'kov formula 
$\ln(T_c/T_{c0}) = \Psi\left(\frac12\right) - \Psi\left(\frac12 + \gamma/2\pi T_c\right)$.
This does not depend on $\delta$ and $\Gamma$ independently, but only on their combination through $\gamma$. 
The critical field, however, does depend on the phase shift $\delta $, and we refer to \cite{VorontsovAB:2008jl} 
for more details of the formulation of this problem. In particular, one can take the limit $\delta \to 0$ explicitly
when evaluating the Born limit.   
Results for magnetization and free energy as function of magnetic field are shown in Fig. \ref{fig:FfromM}. 
Both of these quantities are sensitive to the scattering strength given by the phase shift $\delta$. 

This problem allows us to make an indirect check of the validity of the derived free energy functional. 
The free energy is explicitly given by Eq. \eqref{eq:FE} in terms of self-consistently calculated propagators and order
parameter. This is the free energy in superconducting state compared to the free energy of the normal state in 
the same external field $H$. 

On the other hand, one can calculate the magnetization directly from the impurity-averaged Green's function
\be
\vM_\sm{SC}(T,H) = \vM_\sm{N}(H) 
+ 2\mu N_f \; T \sum_n \langle \vg(\vare_n; \vp, \vR) \rangle \,.
\label{MagSC}
\ee
where the ``high-energy'' normal state response is $M_N(H) = 2\mu^2 N_f \, H$. 
Since magnetization is a
derivative of the free energy, we can reverse this relation and 
calculate the free energy by integrating the magnetization at fixed temperature, 
\be
\Del \Omega(T,H) = \Del \Omega(T,0) - 
\int_0^H \Big[ M_\sm{SC}(T,H') - M_\sm{N}(H')\Big] dH'  \,,
\label{eq:FfromM}
\ee
starting with the zero-field value of the difference $\Del \Omega(T,0)$ that we adjust by hand for this test. 
As shown in Fig. \ref{fig:FfromM}(b) the two expressions for the free energy agree within numerical accuracy, 
and show a jump in the slope (leading to a jump in the magnetization) at the first order transition. 
For the same scattering rate, the critical field in Born limit is lower than that in the unitarity limit, consistent with the result in \cite{VorontsovAB:2008jl}
and previous works. 
Note that near the SC-N transition the calculation of the free energy functional requires more 
angles, more Matsubara energies and more precision to obtain the same numerical accuracy. 

%~~~~~~~~~~~~~~~~~~~~~~~~~~~~~~~~~~~~~~~~~~~~~~~~~~~~~~~~~~~~~~~~~
%~~~~~~~~~~~~~~~~~~~~~~~~~~~~~~~~~~~~~~~~~~~~~~~~~~~~~~~~~~~~~~~~~

\subsection{Superconductor with a domain wall}
\label{sec:dwsc}

%%%%%%%%%%%%%%%%%%%%%%%%%%%%%%%%%%%%%%%%%%%%%%%%%%%%%%%%%%%%%%%%%%%%%%%%%%%%%%%%%
\begin{figure*}[t]
\includegraphics[width = 0.45\linewidth]{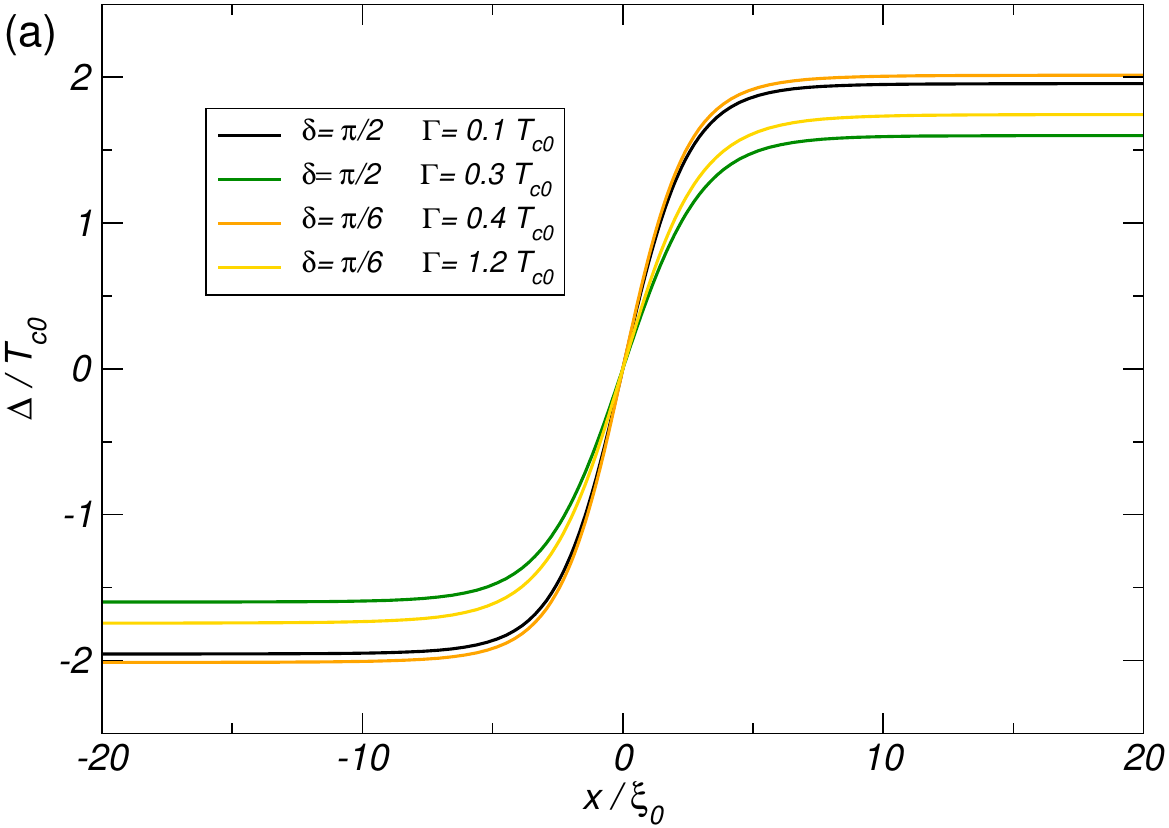}
\hfill
\includegraphics[width = 0.45\linewidth]{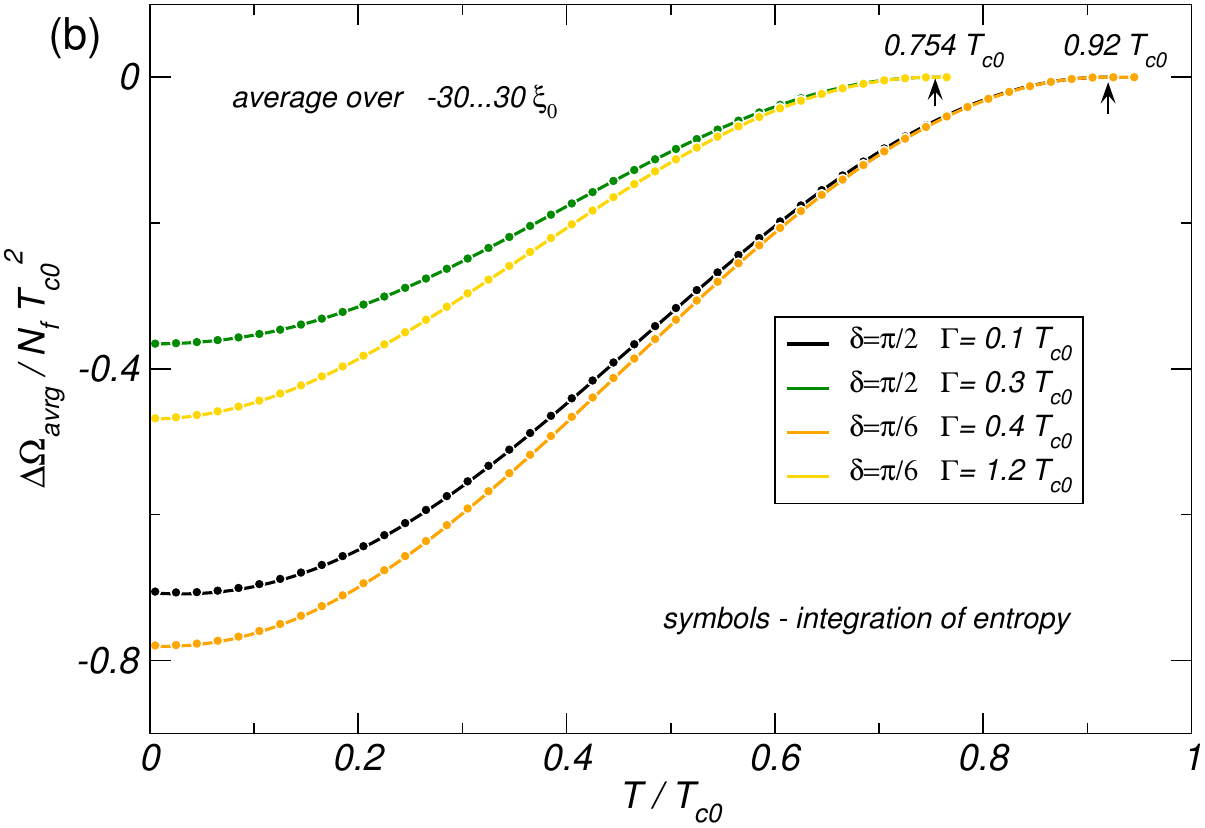}
\caption{ \label{fig:avrgF} 
(a) Domain wall in the $d$-wave superconductor with opposite-sign amplitudes in two domains; 
(b) spatially averaged free energy across the domain wall, over $-30\xi_0 \dots 30\xi_0$.
The coherence length is defined as $\xi_0 = \hbar v_\sm{F}/2\pi k_\sm{B} T_{c0}$. 
We use two values of the scattering rate $\gamma = \Gamma \sin^2\delta = \{0.1, \; 0.3\} T_{c0}$ 
that give the indicated bulk transition temperatures. 
The unitary ($\delta=\pi/2$) and Born ($\delta=\pi/6$) impurities result in different order parameter values, and
different free energies.  
The dot-symbols in (b) are obtained by integration of entropy. %, Eq. \eqref{eq:FfromS}. 
%$N_f$ is the Fermi level DOS for one spin projection. 
%\AV{note for myself - need (a), (b) and change $\Del F$ for $\Del \Omega$}
}
\end{figure*}
%%%%%%%%%%%%%%%%%%%%%%%%%%%%%%%%%%%%%%%%%%%%%%%%%%%%%%%%%%%%%%%%%%%%%%%%%%%%%%%%%

As a second example of a disordered $d$-wave superconductor we look at a non-uniform order parameter, 
that forms a domain wall, where the amplitude 
varies along $x$, $\Delta(\vR) = \Delta(x)$, reaching a uniform state with $\pm \Delta$ far away from $x=0$. 
The profile of the order parameter amplitude for several impurity parameters is shown in
Fig.~\ref{fig:avrgF}(a).
The half-space limit of this configuration is similar to the suppression of superconductivity near a specularly reflecting
interface. 
Although in this example we set the magnetic field to zero, such domain wall configurations are the 
entry-points for the formation of exotic finite center-of-mass momentum FFLO states \cite{VorontsovFFLO} 
that likely have recently been observed in several layered organic superconductors. \cite{Imajo2022,Wosnitza2018}

The total, spatially-integrated, free energy is shown in Fig. \ref{fig:avrgF}(b) for the same impurity parameters. 
We see that for the same scattering parameter $\gamma$ the order parameter and free energy are generally different 
in Born and unitary limits, with Born scatterers corresponding to lower energy. 

It is again interesting to check if our expression for the free energy is consistent with other thermodynamic
observables. Given that the temperature derivative of the free energy is entropy, 
we use an independent way to calculate the entropy of the system. 
From the statistical definition of entropy in a fermionic system in terms of the Fermi distribution function 
$f(\epsilon) = [\exp(\epsilon/T)+1]^{-1}$
and the local density of states,  
\be
S(T,\vR) = -\int\limits_{-\infty}^{+\infty} d\epsilon 
\left[f \ln f +  (1-f) \ln(1-f) \right] 
N(\epsilon, \vR)  \;,
\label{ENTROPY}
\ee
we can obtain the free energy relative to the normal state by integrating the difference between superconducting and
normal states entropies over temperature 
\begin{align}\begin{split}
\Del \Omega(T,\vR) & = \Omega_\sm{SC}(T,\vR) - \Omega_\sm{N}(T) = 
\\
& = -\int\limits_{T_{c}}^T \Big[ S_\sm{SC}(T',\vR) - S_\sm{N}(T') \Big] dT'
\label{eq:FfromS}
\end{split} \end{align}
where $S_\sm{N}(T) = \gamma_s T$ with the Sommerfeld coefficient $\gamma_s = \frac23 N_f \pi^2$. 

The total free energy, integrated over space $\Del \Omega_{avrg}(T) = \int d^3R \; \Del\Omega(T,\vR)$, 
is independent of the calculation method, as is demonstrated in Fig. \ref{fig:avrgF}(b). 
The solid lines correspond to integrated density from Eq.\eqref{eq:FE}, and the dot-symbols to integrated density
from Eq.\eqref{eq:FfromS}. 
The same applies to spatially-uniform systems. 

\begin{figure}[t]
%\includegraphics[width = 0.45\linewidth]{Fig3a.pdf}
%\hfill
\includegraphics[width = 0.99\linewidth]{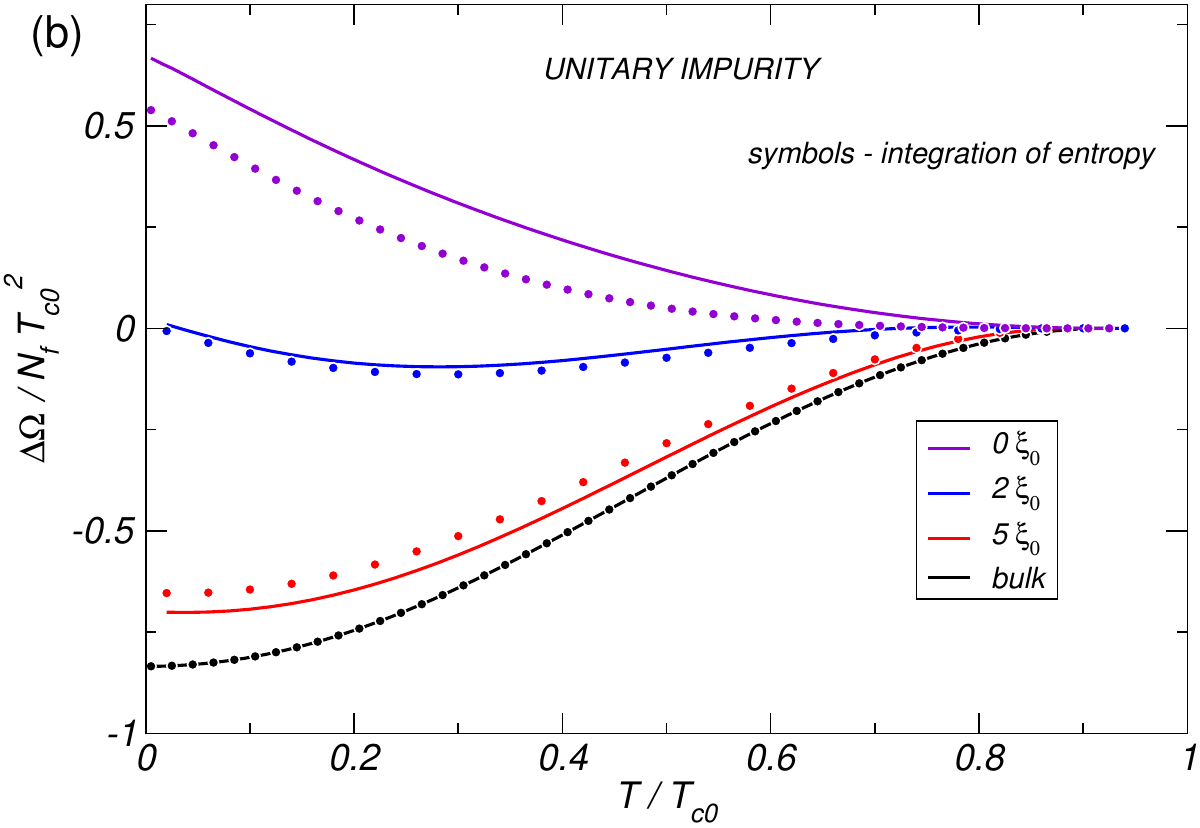}
\caption{ \label{fig:densF} 
The ``spatial density'' of free energy Eq. \eqref{eq:FE} at different locations across the domain wall, shown by the solid lines, 
and ``spatial density'' obtained by the temperature-integrated
entropy Eq. \eqref{eq:FfromS}, shown by dot-symbols. 
``Spatial densities'' obtained in different ways don't agree with each other, except in the uniform (bulk) case, but
generally show the same patterns. 
%which confirms that the spatially-resolved free energy is not necessarily a physically meaningful quantity. 
Shown are unitary impurities; results for the Born limit are very similar. 
}
\end{figure}

However, as we show in Fig. \ref{fig:densF}, the {\it densities} of free energy, 
the one obtained directly from Eq.~\eqref{eq:FE}, and the other from the entropy density, Eq.~\eqref{eq:FfromS}, 
do not numerically agree in regions where the order parameter is non-uniform.
In this respect, the density of thermodynamic quantities on the scale of microscopic %derived 
spatial variations is not a well-defined physical quantity, 
which is a natural consequence of the strong reliance on full spatial integration in the derivation of the free energy
as mentioned around Eq.~\eqref{func_deriv}. 
This  is intuitively clear in the case of entropy, 
since in non-uniform system a quasiparticle bound-state at some energy $\epsilon$ has 
a spatially-dependent probability distribution on the coherence length $\xi_0$, 
and considering different locations on length-scales smaller than $\xi_0$ does not make much sense. 
The overall behavior of the free energy density, however, show the same tendencies in both cases: 
free energy is the highest in the center of the domain wall where pairbreaking is most pronounced, 
becoming smaller as one goes away from it, 
and the temperature evolutions are similar. 

%%%%%%%%%%%%%%%%%%%%%%%%%%%%%%%%%%%%%%%%%%%%%%%%%%%%%%%%%%%%%%%%%%%%%%%%%%%%%%%%%
%%%%%%%%%%%%%%%%%%%%%%%%%%%%%%%%%%%%%%%%%%%%%%%%%%%%%%%%%%%%%%%%%%%%%%%%%%%%%%%%%

\section{Triplet superfluid with Fermi-liquid corrections near interface}
\label{sec:triplet} 

Our final example is a triplet superfluid \He\ near an impenetrable wall. 
We do not include internal impurities here, 
since liquid \He\ is a typically very pure system, 
but instead consider quasiparticle 
scattering at a hard boundary placed at $z=0$, leading to a suppression of unconventional pairing state. 
The system will be in the semi-infinite space $z>0$, uniform in the $(x,y)$ plane, 
and we consider effects of an external magnetic field, applied perpendicular to the interface ($H \parallel z$) 
and parallel to the interface ($H \parallel x$).  
This geometry represents a scenario of competing pairbreaking mechanisms due to orbital scattering off the interface, 
and magnetic suppression of some of the order parameter components.  

Understanding energy balance and realization of different order parameter configurations in this system is quite
important. 
\He\ is a strongly-interacting liquid that has many features useful for understanding novel superconducting
materials with proposed triplet pairing, 
including topological systems.\cite{Mizushima2016}
By itself, \He\ has recently been the focus of investigations due to its modified properties in confined geometries 
\cite{Heikkinen2021,Heikkinen2025}
and in the presence of aerogel pairbreaking disorder.\cite{Halperin2019,Nguyen2024} 
Moreover, there are proposals to use superfluid \He\ for detection of exotic Majorana modes in the 
vicinity of such interfaces, and probing them using magnetic fields.\cite{Silaev2014,Mizushima2018}

The mean-field triplet order parameter is parametrized by a vector in spin space $\vDelta$, and in 4-dimensional
Nambu space is given by  
\be
\whDelta^{mf} = \begin{pmatrix} 0 & (\vDelta \cdot \vsigma) i\sigma_2 \\ i\sigma_2 (\vDelta^* \cdot \vsigma) \end{pmatrix} 
		\equiv \begin{pmatrix} 0 & \Delta^{mf} \\ \tilde \Delta^{mf} & 0 \end{pmatrix}
\ee
where the spin vector depends on the momentum with decomposition in $p$-wave ($\ell=1$) orbitals 
$\{\hp_x,\, \hp_y, \, \hp_z\}$,
\begin{align}
\begin{split}
\Delta_\alpha(\vp,\vR) = \sum_{i=x,y,z} A_{\alpha i} (\vR) \, \hat p_i
\,,\quad \\
A_{\alpha i} =\begin{pmatrix}  A_{xx} & 0 & 0 \\ 0 & A_{yy} & 0 \\ 0 & 0 & A_{zz} \\ \end{pmatrix} 
\end{split}
\end{align}
and in the second line we assumed a particular form of the order parameter matrix 
corresponding to the general B-phase structure, with spin space coordinates aligned with the orbital coordinates.

%~~~~~~~~~~~~~~~~~~~~~~~~~~~~~~~~~~~~~~~~~~~~~~~~~~~~~~~~~~~~~~~~~
\begin{figure}[t!]
\centerline{\includegraphics[trim={0 30 0 60},clip,width=1.1\linewidth]{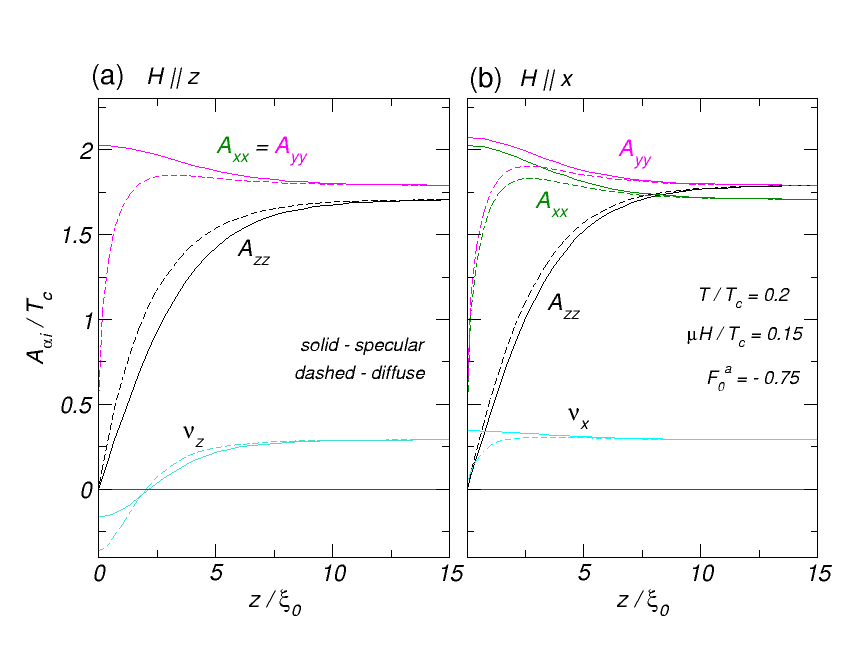}}
\caption{ \label{fig:OPHe3}
The order parameter of a triplet $p$-wave superfluid near a quasiparticle-reflecting surface, in the presence of a magnetic field. 
The quasiparticle scattering from the surface is modeled as specular (solid lines) or diffuse (dashed lines).  
(a) the magnetic field is oriented perpendicular to the surface $H\parallel z$; 
(b) the magnetic field is parallel to the surface, in this case $H \parallel x$. 
}
\end{figure}
%~~~~~~~~~~~~~~~~~~~~~~~~~~~~~~~~~~~~~~~~~~~~~~~~~~~~~~~~~~~~~~~~~
%~~~~~~~~~~~~~~~~~~~~~~~~~~~~~~~~~~~~~~~~~~~~~~~~~~~~~~~~~~~~~~~~~
\begin{figure}[t!]
\centerline{\includegraphics[trim={50 0 100 0},clip,width=1.0\linewidth]{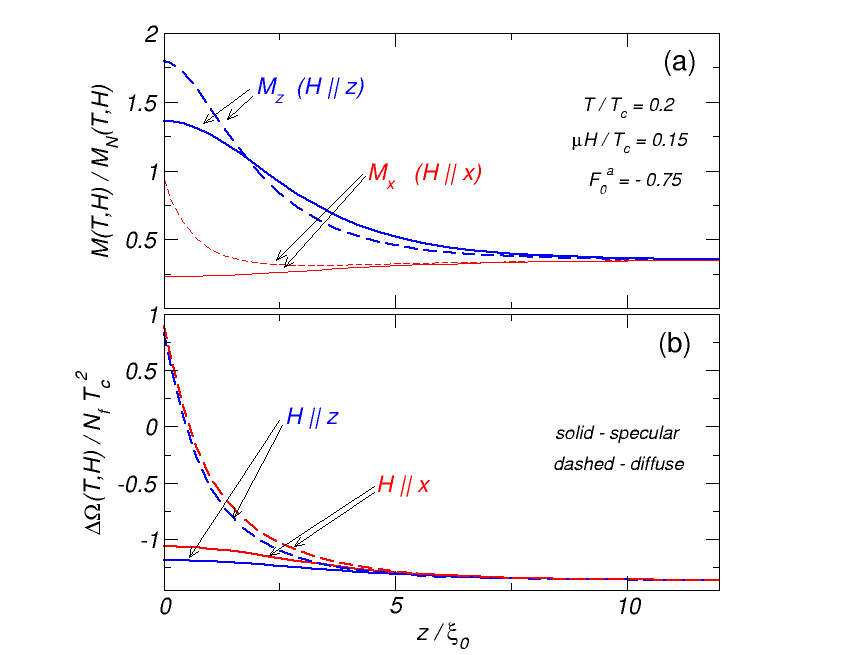}}
\caption{ \label{fig:FEHe3}
(a) Magnetization density near a wall for specular (solid lines) and diffuse (dashed lines) boundary conditions.  
The system behaves as a strongly anisotropic magnetic superfluid, with the magnetization also affected by quasiparticle scattering
through a strongly modified density of states in the two scattering limits.
(b) Free energy ``density'' near a pair-breaking surface in the presence of the external field. 
Diffuse scattering at the boundary is energetically more costly than specular scattering, 
while the perpendicular field configuration, $H \parallel z$ (blue) is energetically
lower than the parallel field configuration, $H \parallel x$ (red). 
}
\end{figure}
%~~~~~~~~~~~~~~~~~~~~~~~~~~~~~~~~~~~~~~~~~~~~~~~~~~~~~~~~~~~~~~~~~

Magnetic field is one of the standard probes to test the spin structure of complex superfluids and superconductors.\cite{min99} 
An external magnetic field couples to the neutron's spin magnetic moment $\mu < 0$ in the nucleus of \He. 
The interaction is given by the
following matrix in Nambu space which includes the `high energy' renormalization of the vertex, 
\be
\widehat U = -(1-A_0^a) \mu \begin{pmatrix} \vsigma \cdot \vH  &  0 \\  0  & \vH \cdot \vsigma^* \end{pmatrix} 
\,.
\ee
The interaction coefficient in the antisymmetric $\ell=0$ molecular field channel for \He\ is 
$$
Z_0^a \equiv 1- A_0^a = \frac{1}{1+F_0^a}\,, \qquad F_0^a \approx -3/4, \quad A_0^a \approx -3  \,,
$$
corresponding to strong ferromagnetic correlations.\cite{vol90}

The magnetic field gives rise to Landau molecular field corrections in the superfluid state, 
which in this case are parametrized by 
\be
\widehat\nu^{mf} = \begin{pmatrix} \vnu \cdot \vsigma &  0 \\  0  & \vnu \cdot \vsigma^* \end{pmatrix} 
		\equiv \begin{pmatrix} \Sigma^{mf} & 0  \\ 0 & \tilde \Sigma^{mf} \end{pmatrix}
\ee
and the spin components of the molecular field $\vnu$ are determined from the self-consistency relation
\be
\nu_\alpha(\vR) = A_0^a \; T\sum_{\vare_m} \langle g_\alpha(i\vare_m; \vp, \vR) \rangle_\sm{FS}  
\ee
where we used the usual quasiclassical notations \cite{Eschrig2009} to parametrize the propagators,  
\be
\widehat g = \begin{pmatrix} 
g & f \\ \tilde f & \tilde g 
\end{pmatrix} 
\ee
where $g, \tilde g, f, \tilde f$ are $2 \times 2$ spin matrices, 
and the molecular field self-consistency relation uses the spin-triplet part $\vg$ of the diagonal term $g = g_0 + \vg \cdot \vsigma$. 

The self-consistent calculations of the non-uniform order parameter in the presence of interface scattering 
using quasiclassical theory follows the standard approaches of Riccati parametrization of the quasiclassical propagators 
and solution of the transport equations.\cite{Eschrig:2000ux}
We compare two limits of the interface scattering of quasiparticles, specular and diffuse, by using 
the random $S$-matrix approach.\cite{Nagato2011}
Results for the various components of the mean-field order parameter and molecular fields are shown in 
Fig. \ref{fig:OPHe3}. 
A pair-breaking surface suppresses the $z$-orbital component of the order parameter $A_{\alpha z}$ and a diffuse surface
suppresses all orbital components. A magnetic field suppresses the $\Delta_\alpha$ component that is along the direction
of the external field, which corresponds to the zero projection component of the spin-triplet state. For $H\parallel z$ this is
the $A_{zz}$ component, while for $H\parallel x$ this is the $A_{xx}$-component. 
The Fermi-liquid molecular field $\vnu$ (shown are only relevant non-zero component) 
changes sign near the surface for $H \parallel z$, enhancing the magnetization. 

We calculate two quantities, magnetization and free energy. 
The Fermi-liquid effects modify expression \eqref{MagSC} for the magnetization,
by including the ``high-energy'' renormalization into the relevant vertices, 
\be
\vM_\sm{SF}(T,H) = \frac{2 \mu^2 N_f}{1+F_0^a} \vH  
+ \frac{2\mu N_f }{1+F_0^a} T \sum_n \langle \vg(\vare_n; \vp, \vR) \rangle_\sm{FS} \,.
\label{MagSF}
\ee
The free energy is calculated using Eq.~\eqref{eq:QCMmodf}, now with both diagonal and off-diagonal mean fields, 
$\widehat\sigma^{mf} = \widehat\nu^{mf} + \whDelta^{mf}$. 
As an example, we explicitly write the expression for the free energy after the trace over the particle-hole space, and using
symmetries between Green's functions with arguments $\vare_m$ and $-\vare_m$,

%%%%
%%%%
%%%%
%%%%
%%%%
\begin{widetext}
\begin{align}
\begin{split} 
\Omega - \Omega_N = \int d^3 R \; N_f \; 2 T \sum_{\vare_m>0} & 
\Re\left[ \int_{\vare_m}^\infty \,  {\rm tr_2} \left\langle  
\frac{ig - i\tilde g}{2} - \pi  - \pi \frac{\Delta^{mf} \tilde\Delta^{mf} }{2\lambda^2} \right\rangle_\sm{FS}  d\lambda  
\right. \\ 
& 
\left. 
- \frac14  {\rm tr_2} \left\langle 
\Sigma^{mf} g + \tilde \Sigma^{mf} \tilde g 
+ \Delta^{mf} \left( \tilde f - \pi \frac{\tilde \Delta^{mf}}{\vare_m} \right) 
+ \tilde \Delta^{mf} \left( f -\pi  \frac{\Delta^{mf}}{\vare_m} \right) 
\right\rangle_\sm{FS}
\right]
\end{split} 
\end{align}
The remaining trace ${\rm tr_2}[\dots]$ is over the spin degrees of freedom. 
We re-iterate that all self energies that appear explicitly in this expression are mean-field, 
but this expression works for the case when the propagator is
self-consistently calculated with impurities as well. 
Mean-field self-consistency equations can be used to reduce the four terms on the second line to just the mean field self-energies  
%products of the kind $|\Delta|^2 \langle |\cY|^2\rangle \ln (T/T_{c})$, 
as was done in Eq.\eqref{eq:FE} (but notice the minus sign for the molecular field term), 
%while molecular fields in a channel $A_\ell^{s,a}$ will appear as $[A^{s,a}_\ell]^{-1} \nu^2$. 
\begin{align}
\begin{split} 
\Omega - \Omega_N = \int d^3 R \; N_f \; \left\{ 
\langle |\vDelta(\vp, \vR) |^2 \rangle_\sm{FS} \ln \frac{T}{T_{c}}  
- \frac{\vnu(\vR)^2}{A^{a}_0}  
+ 2 T \sum_{\vare_m>0} 
\Re \int\limits_{\vare_m}^\infty \,  {\rm tr_2} \left\langle  
\frac{ig - i\tilde g}{2} - \pi  - \pi \frac{\Delta^{mf} \tilde\Delta^{mf} }{2\lambda^2} \right\rangle_\sm{FS}  d\lambda 
\right\} 
\end{split} 
\end{align}
\end{widetext}

The spatial profiles of the observables are shown in Fig. \ref{fig:FEHe3}. 
Magnetization near the interface is enhanced for specular scattering in the case of $H\parallel z$ and suppressed for $H\parallel x$, 
while diffuse scattering leads to an enhancement of both. This agrees with previous calculations for specular \cite{Nagato2009,Mizushima2012a}
and rough surfaces.\cite{Nagato2018}

Although the ``density'' of free energy on the coherence length scale is not a well-defined quantity, it gives  
insight into the relative magnitude of pairbreaking energy costs. Overall, diffuse scattering leads to a significant
energy costs near the interface, associated with reduction in magnitude of all condensate components and large gradient energies. 
In a magnetic field, the energy for the $H\parallel z$ configuration is always lower than the energy for the $H\parallel x$ case. 

We checked that the same total (spatially-integrated) free energy values, that include surface pairbreaking and
Fermi-liquid corrections, can also be obtained by integration of magnetization at fixed $T$, 
as follows from Eq.~\eqref{eq:FfromM}. 

We also note that the competition between magnetic suppression and orbital pairbreaking leads to certain orientation effects 
between spin and orbital spaces. If magnetic field is stronger than the spin-orbit coupling field scale, the spin vector
$\vDelta$ will rotate relative to the orbital direction $\hvp$ to minimize the energy. In case of superfluid \He\ the
scale is given by the weak dipole-dipole energy $H^* \sim 50$G. Above this field the $H \parallel x$ configuration will lead
to a reorientation of the spin-$x$ axis with the orbital-$z$ direction, resulting in an off-diagonal matrix $A_{\alpha i}$.\cite{Mizushima2012a}

%~~~~~~~~~~~~~~~~~~~~~~~~~~~~~~~~~~~~~~~~~~~~~~~~~~~~~~~~~~~~~~~~~

\section{Conclusions}
\label{sec:conclusions} 

In summary, we have derived a new formulation of the thermodynamic potential that gives the free energy of a superconducting 
state in the presence of disorder, such as impurities, and other pairbreaking scattering, for example from interfaces. 
It unifies several previously used approaches and free energy expressions appearing in the literature, and in particular
connects the spectral properties of the system (real energy axis) with Matsubara imaginary energies without performing
numerical analytic continuation, but rather using a flexible contour integration in the complex energy plane. 

This approach allows direct transition from the full propagators formulation to the low-energy quasiclassical propagators. 
In the important case of $t$-matrix impurity scattering we find that the impurity self-energies do not appear in the
free energy thermodynamic potential explicitly, but only through self-consistent calculation of propagators, impurity and mean
field self-energies. 
A particularly compact and simple expression for the free energy is further obtained for the case of energy-independent mean
fields, and we present details of usage of this expression in several examples of disordered non-uniform singlet and triplet
superconductors. 

The more general expression can serve as a convenient starting point for extensions to many other systems and other forms
of disorder. For example, \He\ imbibed into aerogel that introduces disorder, \cite{Halperin2019} is a system with many novel
characteristics; our free energy functional can be naturally used to study this system.

%~~~~~~~~~~~~~~~~~~~~~~~~~~~~~~~~~~~~~~~~~~~~~~~~~~~~~~~~~~~~~~~~~

\section*{Data availability}

The data are not publicly available. The data are available from the authors upon reasonable request. 

\section*{Acknowledgements}
M.~E. acknowledges funding by Deutsche Forschungsgemeinschaft (DFG, German Research Foundation) under Project No. 530670387.
A.~B.~V. acknowledges funding from the U.S. National Science Foundation under DMR-2023928. 

%~~~~~~~~~~~~~~~~~~~~~~~~~~~~~~~~~~~~~~~~~~~~~~~~~~~~~~~~~~~~~~~~~

%\include{GL_section}

%\newpage
\begin{widetext}
\appendix
\section{Properties of the $\otimes$-product} 
\label{OTIMES}

The convolution product is defined in coordinate representation as 
\bea
[\widehat A\otimes \widehat B](\vr_1,\vr_2) = \int d^3 \vr' \; \widehat A(\vr_1,\vr') \widehat B(\vr', \vr_2)  \,.
\label{otimesR}
\eea
For a pair of coordinates $\vr_1,\vr_2$ one defines center-of-mass $\vR = (\vr_1+\vr_2)/2$ and relative
$\vr=\vr_1-\vr_2$ coordinates. 
Wigner representation corresponds to a Fourier transform $\vr \to \vp$. 
In Wigner representation the convolution product is 
\begin{eqnarray}
[\widehat A\otimes \widehat B](\vp,\vR) = e^{-\frac{i}{2} (\nabla^A_{\vp}\nabla^B_{\vR}-\nabla^B_{\vp}\nabla^A_{\vR})} 
\widehat A(\vp,\vR) \widehat B(\vp,\vR) 
\label{otimesW}
\end{eqnarray}
In either coordinate \eqref{otimesR} or Wigner \eqref{otimesW} representation one can prove that the associative property
holds
\be
\widehat A \otimes [\widehat B \otimes \widehat C] = 
[ \widehat A \otimes \widehat B] \otimes \widehat C \qquad  =  
\widehat A \otimes \widehat B \otimes \widehat C 
\label{otimesAssoc}
\ee

Using \eqref{otimesR} it is also straightforward to show that under the trace we can use cyclic permutations,
which follows from the trace property of matrix multiplication:  
\begin{eqnarray}
\int \frac{d^3\vp}{(2\pi)^3} \; [\widehat A \otimes \widehat B](\vp,\vR)
= [\widehat A \otimes \widehat B](\vr=0,\vR) = \int d^3 \vr' \; \widehat A(\vr_1,\vr') \widehat B(\vr', \vr_2)
\Big|_{\vr_1=\vr_2 = \vR}  
\\
\qRq 
{\rm Tr}_\otimes \{ \widehat A \otimes \widehat B \} = 
\int d^3\vR \int \frac{d^3\vp}{(2\pi)^3}
\; {\rm tr}_4 \{ [\widehat A \otimes \widehat B](\vp,\vR) \}
= {\rm Tr}_\otimes \{ \widehat B \otimes \widehat A \} 
\end{eqnarray}
where we used that $\int d^3\vR \int d^3 \vr' \; {\rm tr}_4 \{\widehat A(\vr,\vr') \widehat B(\vr', \vr) \}\big|_{\vr = \vR} =\int d^3\vr \int d^3 \vR' \; {\rm tr}_4 \{\widehat B(\vr',\vr) \widehat A(\vr, \vr') \}\big|_{\vr' = \vR'}$.

\section{Operator Algebra}
\label{OA}

Assume that there are no eigenvalues $\lambda(\widehat A )$ of $\widehat A$ on the negative real axis. 
Then
there exist a uniquely defined principal logarithm $\widehat X=\ln(\widehat A)$ such that $e^{\widehat X}=\widehat A$ and $-\pi < \mbox{Im}[\lambda (\widehat X )] < \pi$.

%Furthermore, there exists for any integer $p>0$ a $p^{\rm th}$ root $\widehat X=\widehat A^{\frac{1}{p}} $ such that $\widehat X^p=\widehat A$ and $-\pi/p < \mbox{arg}[\lambda(\widehat X )] < \pi/p$.

Furthermore there is an integral formula for the operator logarithm given by
\begin{equation}
\ln(\widehat A+\widehat D )-\ln(\widehat A) = \int\limits_0^\infty dt \left(
[\widehat A +t\widehat 1]^{-1} \otimes \widehat D \otimes [\widehat A +t\widehat 1]^{-1} -
[\widehat A +t\widehat 1]^{-1} \otimes \widehat D \otimes [\widehat A +t\widehat 1]^{-1} \otimes \widehat D \otimes [\widehat A +t\widehat 1]^{-1} 
+ \cdots \right) 
\end{equation}
%which also can be used to define a generalized derivative (``Frech\'et derivative'').
%We denote the trace $\int d^3\vR \int d^3 \vp/(2\pi)^3 \mbox{tr}_4\{\cdots\}$ by $\mbox{Tr}_\otimes \{\cdots\}$.
This leads directly (using cyclic invariance of trace) to
\begin{eqnarray}
\mbox{Tr}_\otimes \left\{ \ln(\widehat A+\widehat D )-\ln(\widehat A) \right\}&= &
\mbox{Tr}_\otimes  \left\{ \int_0^\infty dt [\widehat A +t\widehat 1]^{-2} \otimes \widehat D \right\}
+ {\cal O}(||\widehat D||^2)  
=\mbox{Tr}_\otimes [ \widehat A^{-1}\otimes \widehat D ] + {\cal O}(||\widehat D||^2) .
\end{eqnarray}
Note that Tr$_\otimes [\widehat A^{-1}\otimes \widehat D ]$=Tr$_\otimes [\widehat D \otimes \widehat A^{-1}]$.
In particular, this gives for setting $\widehat D= d\widehat A(z)$ and dividing by $dz$
\begin{equation}
\frac{d}{dz} \ln(\widehat A(z) ) = \int_0^\infty dt 
[\widehat A +t\widehat 1]^{-1} \otimes \frac{d\widehat A }{dz} \otimes [\widehat A +t\widehat 1]^{-1} 
\end{equation}
or under the trace
\begin{eqnarray}
\mbox{Tr}_\otimes \left\{ \frac{d}{dz} \ln(\widehat A(z) )\right\}
=\mbox{Tr}_\otimes  \left\{ \widehat A^{-1}\otimes \frac{d\widehat A }{dz}\right\}
\end{eqnarray}

\section{Large-energy asymptotic behavior of propagators} 
\label{ASYMPTG}

When discussing $\xi$- and $z$- integrations, it is important to know the asymptotic behavior
of the Green's function, and we record it here.
The key element of this analysis are the commutation relations between  
$\widehat\tau_3$, off-diagonal $\whDelta \propto \widehat\tau_{1,2}$ and $\widehat \nu, \widehat U \propto \widehat 1, \widehat\tau_3$.
%($\widehat 1$ component of $\widehat \nu $ can be absorbed into unit term of 
%$\whG_0^{-1} = z\widehat\tau_3 - \xi$) 
Assuming large $|z|$ in $\whG_0^{-1} = \widehat\tau_3 z - \xi_\vp$ we  
expand $\whG = [\whG_0^{-1} - \widehat U - \widehat \Sigma]^{-1}$:  
\be
\whG  = \whG_0 + \whG_0 \otimes (\widehat U + \widehat \Sigma) \otimes \whG_0
+ \whG_0 \otimes (\widehat U + \widehat \Sigma) \otimes \whG_0 \otimes (\widehat U + \widehat \Sigma) \otimes \whG_0 + \dots 
\ee
Each $\grad_\vp$ gradient in the convolution product with $\whG_0(\vp)$ will create an additional power of $\whG_0$, since 
$\grad_\vp \whG_0(\vp) = (\widehat\tau_3 z - \xi_\vp)^{-2} \, \vv_f(\vp) = \whG_0^2 \, \vv_f(\vp)$. 
Therefore, if we consider only terms to order $\whG_0^3$, we can replace the convolution products in the last term by
usual matrix products. In the second term we use 
\[
\whG_0(\vp) \otimes \widehat F(\vp,\vR) \otimes \whG_0(\vp) =  
\whG_0 \widehat F \whG_0 
- {i\over2} \whG_0^2 (\vv_f\cdot\grad_\vR \widehat F) \whG_0  
+ {i\over2} \whG_0 (\vv_f\cdot\grad_\vR \widehat F) \whG_0^2 ~~~ + \cO(\whG_0^4) 
\]
to obtain the asymptotic expansion 
\begin{align} \begin{split} 
\label{asymptG}
\whG  & 
= \frac{1}{ z\widehat\tau_3 - \xi} 
+ \frac{\widehat\nu +\widehat U }{ (z\widehat\tau_3 - \xi)^2} 
+ \frac{\whDelta }{ \xi^2 - z^2 } 
+ i \frac{z\widehat\tau_3}{(\xi^2 - z^2)^2} (\vv_f \cdot \grad_\vR \whDelta) 
+ \frac{(\widehat U+ \widehat\nu)^2 }{ (z\widehat\tau_3 - \xi)^3} 
\\ & 
+ \whDelta^2 \frac{1}{ (\xi^2 - z^2)(z\widehat\tau_3 - \xi) } 
+ \whDelta(\widehat U+ \widehat\nu) \frac{1}{ (\xi^2 - z^2)(z\widehat\tau_3 - \xi) } 
+ \frac{1}{ (\xi^2 - z^2)(z\widehat\tau_3 - \xi) } (\widehat U+ \widehat\nu)\whDelta 
+ \dots  \;.
\end{split} \end{align} 
We can use this to find the asymptotic form of the quasiclassical propagator as well, by $\xi$-integrating: 
\begin{align} \begin{split} 
\label{asymptQCG}
\lim_{|z|\to\infty} \whg(z)  & = \int\limits_{-\infty}^{+\infty} d\xi \; \whG(z,\xi) 
\\ & 
=  -i \pi \, \sgn(\Im z) \; \widehat\tau_3 
+  i \pi \, \sgn(\Im z) \; \frac{\whDelta }{z} 
+  i \pi \, \sgn(\Im z) \; \frac{1}{2 z^2} \widehat\tau_3 
\left[-i \vv_f \cdot \grad_\vR \whDelta +\whDelta^2 + (\widehat U+ \widehat\nu)\whDelta -  \whDelta (\widehat U+ \widehat\nu) \right] 
+ \dots 
\end{split} \end{align} 

The first term in \eqref{asymptQCG} is the normal state low-energy propagator, coming from the branch-cut in $\ln z =
\int d\xi/(z-\xi)$; 
the terms with poles on one side of the complex $\xi$-plane disappear, such as $1/(\widehat\tau_3 z - \xi)^2$, 
and terms that involve poles on both side of complex $\xi$-plane, such as $1/(\xi^2-z^2)$, contribute to terms with
off-diagonal self-energy $\whDelta$. 
Note that $\widehat \Delta$ and $\widehat \nu$ are in general themselves functions of $z$, which however do not diverge
for large $|z|$ and merely give rise to additional higher-order contributions in an asymptotic expansion for large $|z|$.

\section{Coupling to electromagnetic fields} 
\label{EM}

As discussed in the main text, the full thermodynamic potential is 
a sum of two contributions, $\Omega_{\rm tot}=\Omega+\Omega_{\rm EM}.$  
The fields $(\phi,\vA)$ couple to the matter through gauge-invariant terms in $\Omega$, and have separate energy
functional $\Omega_{\rm EM}[\phi,\vA]$.   
The electromagnetic vector potential $\vA$ enters the microscopic Hamiltonian by replacing the canonical momenta by
their kinetic equivalents. However, within our expansion the vector potential it is of order $s$ (producing magnetic
fields maximally of the order of the superconducting critical magnetic fields) and varies on the scale of the
superconducting coherence length (such that spatial derivatives of the vector potential add additional orders of $s$).
Therefore, it enters within our theory as part of the the potential $\widehat U$, specifically in the form
$-{e\over c} \widehat\tau_3\vv_f(\vp_f)\cdot \vA (\vR)$ with the Fermi velocity $\vv_f(\vp_f)$ and the electronic charge $e<0$
\cite{Alexander1985a}. 
Note, however,
that the Green's functions and self energies in general depend non-linearly on $\vA$. This implementation of the
electromagnetic vector potential allows the use of the convolution product in Wigner representation in terms of
canonical momenta. 
\\
%As discussed in the main text, the electromagnetic vector potential enters within our theory as part of the external
%potential $\widehat U$, specifically in the form $-(e/c)\widehat\tau_3\vv_f(\vp_f)\cdot \vA (\vR)$ with the Fermi velocity
%$\vv_f(\vp_f)$ and the electronic charge $e<0$ \cite{Alexander1985a}. 

The scalar potential $\phi$ enters  $\widehat U$
in the form $e  \Zc(\vp_f) \phi (\vR) \widehat 1$ with the high-energy Landau renormalization factor $\Zc(\vp_f)$. 
In metals, charge fluctuations due to quasiparticles are very effectively screened by high-energy electrons on the scale
of the Thomas-Fermi screening length. For each charge pileup due to quasiparticles, a corresponding charge depletion of
high-energy electrons leads to local charge neutrality when averaged over a sphere of diameter large compared to the
Thomas-Fermi screening length. As a result, the scalar potential fulfills the condition of local charge neutrality,
which is valid in leading order in $s$ \cite{GorkovKopnin75, ArtemenkoVolkov79, Eschrig2009b}. Each quasiparticle charge
fluctuation is almost instantaneously (on a time scale of the inverse plasma frequency) compensated by high-energy
electrons. The scalar potential $\phi $ combines with the chemical potential $\mu$ to yield the electrochemical
potential $\mu-e\phi$, the variation of which determines the voltage between different spatial points. A consequence of
all this is that due to the absence of an induced charge density on low-energy length scales, there is no electric field
energy density contribution to the thermodynamic potential in leading order in $s$.

Finally, the Zeeman coupling enters in the form 
$-\mu_e \widehat \vS \cdot \Zs (\vp_f) \cdot \vB(\vR)$, 
with 
$\widehat \vS={\rm diag}(\vsigma,\vsigma^\ast)$ 
where $\vsigma $ is the vector of spin Pauli matrices, and the tensor
$\Zs(\vp_f)$ renormalizes the electron magnetic moment $\mu_e<0$ due to
high-energy electronic scattering processes. 

Therefore, the electromagnetic contributions lead to corresponding terms in
$\widehat U$ according to 
\bea 
\widehat U[\phi, \vA] & \equiv &  
\widehat U_{\rm el} [\phi] + \widehat U_{\rm orb}[\vA]+\widehat U_{\rm spin}[\vA]+\cdots 
\nonumber \\ & = & 
% SI 
%e  \Zc(\vp_f) \phi(\vR)  \widehat 1  - e \vv_f(\vp_f) \cdot \vA(\vR)\widehat\tau_3  
%- \mu_e \widehat \vS \cdot \Zs(\vp_f) \cdot \nabla \times \vA(\vR) + \cdots 
% Gaussian
e  \Zc(\vp_f) \phi(\vR)  \widehat 1  - {e\over c} \vv_f(\vp_f) \cdot \vA(\vR)\widehat\tau_3  
- \mu_e \widehat \vS \cdot \Zs(\vp_f) \cdot \nabla \times \vA(\vR) + \cdots 
\eea 
Variations of the material functional yield
\be
\frac{\delta \Omega}{\delta \phi(\vR)}=\rho_{\rm ind}(\vR), % SI=Gauss
\qquad
%SI
%\frac{\delta \Omega}{\delta \vA(\vR)}=-\vJ_{\rm orb}(\vR)-\nabla \times \vM_{\rm spin}(\vR)
%Gauss
\frac{\delta \Omega}{\delta \vA(\vR)}=- \frac1c \vJ_{\rm orb}(\vR)-\nabla \times \vM_{\rm spin}(\vR)
\ee
with source terms
\bea
\rho_{\rm ind}(\vR) &=&  \frac{e N_f}{2} T \sum_{\varepsilon_n}{\rm tr}_4\langle \Zc(\vp_f) \widehat g(i\varepsilon_n,\vp_f,\vR)\rangle_\sm{FS} -2e^2 N_f \langle  \Zc(\vp_f)\rangle_\sm{FS} \, \phi(\vR) 
\label{sourceRHO}
\\
\vJ_{\rm orb}(\vR)&=& \frac{e N_f}{2} T \sum_{\varepsilon_n}{\rm tr}_4\langle \widehat\tau_3 \vv_f(\vp_f) \widehat g(i\varepsilon_n,\vp_f,\vR)\rangle_\sm{FS}
\label{sourceJ}
\\
\vM_{\rm spin}(\vR)&=& \frac{e N_f}{2} T \sum_{\varepsilon_n}{\rm tr}_4\langle \widehat{\vS}\cdot \Zs(\vp_f) \widehat g(i\varepsilon_n,\vp_f,\vR)\rangle_\sm{FS} +2\mu_e^2 N_f \langle  \Zs(\vp_f)\rangle_\sm{FS} \cdot \vB(\vR) 
\label{sourceM}
\eea
where the last terms in \eqref{sourceRHO} and \eqref{sourceM} are coming from the finite high-energy susceptibility term $\Omega_0^{(2)}[\widehat U]$ in the free
energy. In equation \eqref{sourceM}, a possible equilibrium spin magnetization $\vM_0(\vR)$, which would result from  $\Omega_0^{(1)}[\widehat U]$, should be added when dealing with itinerant ferromagnets; a corresponding term $\rho_0(\vR)$ in \eqref{sourceRHO} is, however, absent in metals as it is canceled by the ionic background.
The Fermi surface average is weighted by the angle-resolved density of states $N(\vp_f)$ at each Fermi surface point $\vp_f$. The quantity $\langle  \Zc(\vp_f)\rangle_\sm{FS}$ is often denoted as $Z_0^s$ and the quantity $\langle  \Zs(\vp_f)\rangle_\sm{FS}$, if diagonal, as $Z_0^a \overset{\scriptscriptstyle\leftrightarrow}{1}$. 
The condition of local charge neutrality reads $\rho_{\rm ind}(\vR)=0$; in equilibrium, this leads to $\phi(\vR)=0$.
 The orbital current $\vJ_{\rm orb}$ and the spin magnetization $\vM_{\rm spin}$ can be obtained separately if in the
Zeeman term $\nabla \times \vA$ is replaced by $\vB_{\rm spin}$, and the variations are performed with respect to $\vA$
and $\vB_{\rm spin}$ separately. At the end $\vB_{\rm spin}=\nabla \times \vA$ has to be invoked. Then
\be
\left(\frac{\delta \Omega}{\delta
%SI
%\vA(\vR)}\right)_{\vB_{\rm spin}}=-\vJ_{\rm orb}(\vR), \qquad
% Gauss
\vA(\vR)}\right)_{\vB_{\rm spin}}=-\frac1c \vJ_{\rm orb}(\vR), \qquad
\left(\frac{\delta \Omega}{\delta \vB_{\rm spin}(\vR)}\right)_{\vA}=-\vM_{\rm spin}(\vR). %SI = Gauss
\ee

The magnetic field energy density contribution to the thermodynamic %Kadanoff-Baym 
functional is
\be
% SI
%\Omega_{\rm EM}[\phi, \vA]=\int d^3\vR \left[ - \frac{\varepsilon_0(-\nabla \phi-\vE_a)^2}{2}+ \frac{(\nabla \times \vA-\vB_a)^2}{2\mu_0}\right]
% Gaussian
\Omega_{\rm EM}[\vA] =\int d^3\vR \left[  \frac{(\nabla \times \vA-\vB_a)^2}{8\pi}\right]
\ee
where 
%$\vE_a(\vR)$ and 
$\vB_a(\vR)$ is a (fixed) external field.
Variation of this term gives
\be
%SI
%\frac{\delta \Omega_{\rm EM}}{\delta \phi(\vR)}=-\varepsilon_0 \left[-\nabla^2 \phi(\vR)-\nabla \vE_a(\vR)\right],
% Gauss
\frac{\delta \Omega_{\rm EM}}{\delta\vA(\vR)}
%SI
% =\frac{1}{\mu_0}\left[\nabla \times \Big( \nabla \times \vA(\vR)\Big)-\nabla \times \vB_a(\vR)\right].
%Gauss
=\frac{1}{4\pi}\left[\nabla \times \Big( \nabla \times \vA(\vR)\Big)-\nabla \times \vB_a(\vR)\right].
\ee
Variation of $\Omega_{\rm tot}$ with respect to the vector potentials at the stationary point yields 
%The variation of the total functional gives
\bea
\frac{\delta \Omega_{\rm tot}}{\delta \vA(\vR)}=0 \quad \Longrightarrow \quad \nabla \times \Big(\vB(\vR)-\vB_a(\vR)-4\pi \vM_{\rm spin}(\vR)\Big)=\frac{4\pi}{c} \vJ_{\rm orb}(\vR) .
\eea
For evaluating $\Omega_{\rm EM}$ at the extremal solution of $\Omega_{\rm tot}$, the relation
\be
%SI
%\Omega_{\rm EM}=-\frac{1}{2}\int d^3\vR \Big( ( \phi-\phi_a )\rho_{\rm ind}  + ( \vA - \vA_a ) \cdot \vJ_{\rm orb}\Big)
% Gauss
\Omega_{\rm EM}=-\frac{1}{2}\int d^3\vR \left[ \frac1c ( \vA - \vA_a ) \cdot \vJ_{\rm orb}\right]
\ee
with $\nabla \times \vA_a=\vB_a$ is useful, as the integral only extends over the superconductor. 
%Finally, the condition of local charge neutrality gives
%\be
%e\phi(\vR) = \frac{T \sum_{\varepsilon_n}{\rm tr}_4\langle \Zc(\vp_f) \widehat g(i\varepsilon_n,\vp_f,\vR)\rangle_\sm{FS}}{4\langle  \Zc(\vp_f)\rangle_\sm{FS}} .
%\ee
\end{widetext}

%~~~~~~~~~~~~~~~~~~~~~~~~~~~~~~~~~~~~~~~~~~~~~~~~~~~~~~~~~~~~~~~~~~~~~~~~~~~~~
%~~~~~~~~~~~~~~~~~~~~~~~~~~~~~~~~~~~~~~~~~~~~~~~~~~~~~~~~~~~~~~~~~~~~~~~~~~~~~
\bibliographystyle{apsrev4-2}
\bibliography{bibFE,Books}
%\bibliography{bibZotero,Books}

\end{document}

%% file: feynman_diagrams_mf.tex
%to update the diagrams - remove impurity_feynman.*
%    then compile in 3 steps:
% > pdflatex paper.tex  - takes some time to finish
% > mf '\mode:=laserjet; input meanfield_feynman.mf'
% > pdflatex paper.tex
\begin{fmffile}{meanfield_feynman}
\raisebox{14mm}{$\displaystyle  \Phi^{mf}[\widehat G] \quad = $ \quad }
\raisebox{14mm}{$\displaystyle \frac12 \; $}
\raisebox{5mm}{
\begin{fmfgraph*}(30,20)
%\fmfpen{thick}
\fmfleftn{l}{1}
\fmfrightn{r}{1}
\fmfv{label=$G$,label.dist=-45}{l1}
\fmfv{label=$G$,label.dist=-45}{r1}
%\fmfrpolyn{shaded,label=$\Gamma$}{G}{4}
%\fmfpolyn{empty,tension=0.4,label=$A$}{A}{4} %\fmflabel{$1$}{A1} \fmflabel{$2$}{A2} \fmflabel{$3$}{A3} \fmflabel{$4$}{A4}
\fmfpolyn{shaded,tension=0.4,label=$A$,l.a=90,l.d=60}{A}{4} %\fmflabel{$1$}{A1} \fmflabel{$2$}{A2} \fmflabel{$3$}{A3} \fmflabel{$4$}{A4}
\fmf{fermion,right=.7}{A1,r1}
\fmf{fermion,right=.7}{r1,A2}
\fmf{fermion,left=.7}{A4,l1}
\fmf{fermion,left=.7}{l1,A3}
\end{fmfgraph*}
}
\raisebox{14mm}{$\displaystyle \quad + \frac12 \; $}
\raisebox{5mm}{
\begin{fmfgraph*}(30,20)
%\fmfpen{thick}
\fmfleftn{l}{1}
\fmfrightn{r}{1}
\fmfv{label=$\ul G$,label.dist=-45}{l1}
\fmfv{label=$\ul G$,label.dist=-45}{r1}
%\fmfrpolyn{shaded,label=$\Gamma$}{G}{4}
\fmfpolyn{shaded,tension=0.4,label=$A$,l.a=90,l.d=60}{A}{4} %\fmflabel{$1$}{A1} \fmflabel{$2$}{A2} \fmflabel{$3$}{A3} \fmflabel{$4$}{A4}
\fmf{fermion,right=.7}{A1,r1}
\fmf{fermion,right=.7}{r1,A2}
\fmf{fermion,left=.7}{A4,l1}
\fmf{fermion,left=.7}{l1,A3}
\end{fmfgraph*}
}
\raisebox{14mm}{$\displaystyle \quad + \; $}
\begin{fmfgraph*}(20,30)
%\fmfpen{thick}
\fmftop{t1}
\fmfbottom{b1}
\fmfv{label=$F$,label.dist=-45}{t1}
\fmfv{label=$\ul F$,label.dist=-45}{b1}
%\fmfpolyn{full,tension=0.4,label=\textcolor{white}{$V$}}{V}{4} 
\fmfpolyn{hatched,tension=0.4,label=$V$,l.a=0,l.d=60}{V}{4} 
\fmf{fermion,right=.7}{t1,V3}
\fmf{fermion,left=.7}{t1,V2}
\fmf{fermion,right=.7}{V4,b1}
\fmf{fermion,left=.7}{V1,b1}
\end{fmfgraph*}

\end{fmffile}

%% file: feynman_diagrams_imp.tex
%to update the diagrams - remove impurity_feynman.*
%    then compile in 3 steps:
% > pdflatex paper.tex  - takes some time to finish
% > mf '\mode:=laserjet; input impurity_feynman.mf'
% > pdflatex paper.tex
\begin{fmffile}{impurity_feynman}
\raisebox{3mm}{\; $\Sigma_{imp}(\vp,z)  = \qquad $\;}
\begin{fmfgraph*}(10,12) 
     	\fmfbottom{i1} 
	\fmftop{o1}
	\fmfv{decor.shape=cross,decor.size=2mm}{i1}
	\fmfblob{.2h}{o1}
	\fmfv{decor.shape=circle,decor.filled=gray50,decor.size=2.5mm}{o1} 
	\fmf{dashes,tension=0,label=$\widehat{v}$}{i1,o1}
\end{fmfgraph*}
\raisebox{3mm}{\; $\quad + \quad $ \;}
\begin{fmfgraph*}(12,12)
     	\fmfbottom{i1,i2} 
	\fmftop{o1}
	\fmfv{decor.shape=cross,decor.size=2mm}{i1,i2}
	\fmfv{decor.shape=circle,decor.filled=gray50,decor.size=2.5mm}{o1} 
	\fmf{plain,label=$\widehat{G}$}{i1,i2}
	\fmf{dashes,tension=0}{i1,o1}
	\fmf{dashes,tension=0}{i2,o1}
\end{fmfgraph*}
\raisebox{3mm}{\; $\quad + \quad $ \;}
\begin{fmfgraph*}(17,12)
	\fmfstraight
     	\fmfbottom{i1,i2,i3} 
	\fmftop{o1}
	\fmfv{decor.shape=cross,decor.size=2mm}{i1,i2,i3}
	\fmfv{decor.shape=circle,decor.filled=gray50,decor.size=2.5mm}{o1} 
	\fmf{plain}{i1,i2}
	\fmf{plain}{i2,i3}
	\fmf{dashes,tension=0}{i1,o1}
	\fmf{dashes,tension=0}{i2,o1}
	\fmf{dashes,tension=0}{i3,o1}
\end{fmfgraph*}
\raisebox{3mm}{\; $+ \cdots + $ \;}
\begin{fmfgraph*}(25,12)
	\fmfstraight
     	\fmfbottom{i1,i2,i3,i4} 
	\fmftop{o1}
	\fmfv{decor.shape=cross,decor.size=2mm}{i1,i2,i3,i4}
	\fmfv{decor.shape=circle,decor.filled=gray50,decor.size=2.5mm}{o1} 
	\fmf{plain}{i1,i2}
	\fmf{plain}{i2,i3}
	\fmf{plain}{i3,i4}
	\fmf{dashes,tension=0}{i1,o1}
	\fmf{dots,tension=0,label=$(n)$}{i2,o1}
	\fmf{dots,tension=0}{i3,o1}
	\fmf{dashes,tension=0}{i4,o1}
\end{fmfgraph*}
\raisebox{3mm}{\; $+ \cdots $ \,, \;}

\vspace{6mm}
%This self-energy is generated from the `wheel' diagrams 
%\\

\raisebox{8mm}{$\displaystyle  \Phi_{imp}[\widehat G] \quad = $ \quad }
\begin{fmfgraph*}(20,20) 
	\fmfsurroundn{i}{4} 
	%\fmfdotn{i}{4}
	\fmfpolyn{plain,smooth}{v}{4}
	\fmfv{decor.shape=cross,decor.size=2mm}{v4}
	\fmf{phantom,tension=50}{v1,i1}
	\fmf{phantom,tension=50}{v2,i2}
	\fmf{phantom,tension=50}{v3,i3}
	\fmf{phantom,tension=50}{v4,i4}
	\fmf{phantom,tension=0.01}{v2,o1}
	\fmf{dashes,tension=0.01}{v4,o1}
	\fmfv{decor.shape=circle,decor.filled=gray50,decor.size=2.5mm}{o1} 
\end{fmfgraph*}
\raisebox{8mm}{$\displaystyle \quad + \frac12 \; $}
\begin{fmfgraph*}(20,20) 
	\fmfsurroundn{i}{4} 
	%\fmfdotn{i}{4}
	\fmfpolyn{plain,smooth}{v}{4}
	\fmfv{decor.shape=cross,decor.size=2mm}{v2,v4}
	\fmf{phantom,tension=50}{v1,i1}
	\fmf{phantom,tension=50}{v2,i2}
	\fmf{phantom,tension=50}{v3,i3}
	\fmf{phantom,tension=50}{v4,i4}
	\fmf{dashes,tension=0.01}{v2,o1}
	\fmf{dashes,tension=0.01}{v4,o1}
	\fmfv{decor.shape=circle,decor.filled=gray50,decor.size=2.5mm}{o1} 
\end{fmfgraph*}
\raisebox{8mm}{$\displaystyle \quad + \frac13 \; $}
\begin{fmfgraph*}(20,20) 
	\fmfsurroundn{i}{3} 
	%\fmfdotn{i}{3}
	\fmfpolyn{plain,smooth}{v}{3}
	\fmfvn{decor.shape=cross,decor.size=2mm}{v}{3} 
	\fmf{phantom,tension=18}{v1,i1}
	\fmf{phantom,tension=18}{v2,i2}
	\fmf{phantom,tension=18}{v3,i3}
	\fmf{dashes,tension=0.01}{v1,o1}
	\fmf{dashes,tension=0.01}{v2,o1}
	\fmf{dashes,tension=0.01}{v3,o1}
	\fmfv{decor.shape=circle,decor.filled=gray50,decor.size=2.5mm}{o1} 
\end{fmfgraph*}
\raisebox{8mm}{$\displaystyle + \dots + \frac1n \; $}
\begin{fmfgraph*}(20,20) 
	\fmfsurroundn{i}{4} 
	%\fmfdotn{i}{4}
	\fmfpolyn{plain,smooth}{v}{4}
	\fmfvn{decor.shape=cross,decor.size=2mm}{v}{4}
	\fmf{phantom,tension=50}{v1,i1}
	\fmf{phantom,tension=50}{v2,i2}
	\fmf{phantom,tension=50}{v3,i3}
	\fmf{phantom,tension=50}{v4,i4}
	\fmf{dashes,tension=0.01}{v1,o1}
	\fmf{dots,tension=0.01}{v2,o1}
	\fmf{dots,tension=0.01,label=$(n)$}{v3,o1}
	\fmf{dashes,tension=0.01}{v4,o1}
	\fmfv{decor.shape=circle,decor.filled=gray50,decor.size=2.5mm}{o1} 
\end{fmfgraph*}
\raisebox{8mm}{$\displaystyle + \dots \; $}
\end{fmffile}